\documentclass[superscriptaddress, twocolumn,showkeys,
amssymb,amsmath,nobibnotes,aps,prd,
nofootinbib]{revtex4-2}
\pdfoutput=1
\usepackage{svg}
\usepackage{graphicx,subfigure,bm,color,psfrag,hyperref}
\usepackage[export]{adjustbox}
\usepackage{amsfonts}
\usepackage{lipsum}
\usepackage{mathtools}
\usepackage{verbatim}
\usepackage[normalem]{ulem}
\usepackage[dvipsnames]{xcolor}
\hypersetup{colorlinks,linkcolor={blue},citecolor={red},urlcolor={greenW}}  
\definecolor{greenW}{rgb}{0.0, 0.55, 0.1}

\begin{document}

\title{One-parameter dynamical dark energy: Hints for oscillations}

\author{Daniel A. Kessler}
\email{ph4dke@sheffield.ac.uk}
\affiliation{School of Mathematical and Physical Sciences, University of Sheffield, Hounsfield Road, Sheffield S3 7RH, United Kingdom}

\author{Luis A. Escamilla}
\email{l.a.escamilla@sheffield.ac.uk}
\affiliation{School of Mathematical and Physical Sciences, University of Sheffield, Hounsfield Road, Sheffield S3 7RH, United Kingdom}

\author{Supriya Pan}
\email{supriya.maths@presiuniv.ac.in}
\affiliation{Department of Mathematics, Presidency University, 86/1 College Street, Kolkata 700073, India}
\affiliation{Institute of Systems Science,  Durban University of Technology, Durban 4000, Republic of South Africa}

\author{Eleonora Di Valentino}
\email{e.divalentino@sheffield.ac.uk}
\affiliation{School of Mathematical and Physical Sciences, University of Sheffield, Hounsfield Road, Sheffield S3 7RH, United Kingdom}

%\date{}

\begin{abstract}
There is mounting evidence from multiple cosmological probes that dark energy may be dynamical, with an equation of state that evolves over cosmic time. While this evidence is typically quantified using the Chevallier–Polarski–Linder (CPL) parametrization, based on a linear expansion of $w(a)$ in the scale factor, non-parametric reconstructions frequently suggest non-linear features, particularly at late times. In this work, we investigate four minimal one-parameter models of dark energy with non-linear dependence on the scale factor. These models are constrained using Cosmic Microwave Background (CMB) data from \textit{Planck}, lensing reconstruction from \textit{Planck} and ACT-DR6, Baryon Acoustic Oscillation (BAO) measurements from DESI-DR2, and three Type Ia supernovae (SNe) samples (PantheonPlus, DESY5, and Union3), considered independently. Although our conclusions depend on the choice of SNe sample, we consistently find a preference, as measured by the chi-squared statistic and the Bayesian evidence, for these dynamical dark energy models over the standard $\Lambda$CDM model. Notably, with the PantheonPlus dataset, one model shows strong Bayesian evidence ($\Delta \ln B \simeq 4.5$) over CPL, favoring an equation of state that peaks near $a \simeq 0.7$ and oscillates near the present day. These results highlight the impact of SNe selection and contribute to the growing collection of evidence for late-time deviations from $\Lambda$CDM.

\end{abstract}
%-----------------------------------------------
\maketitle
%-----------------------------------------

\section{Introduction}

The discovery of the late-time accelerated expansion of the universe~\cite{SupernovaSearchTeam:1998fmf,SupernovaCosmologyProject:1998vns} significantly unbalanced our understanding of the physical laws and fundamental fields behind cosmic evolution. In a statistically homogeneous and isotropic universe,
these observations are most easily explained by a hypothetical fluid with negative pressure (referred to as \textit{dark energy}), traditionally characterized by a positive cosmological constant $\Lambda$ with an equation of state $w_{\Lambda} = -1$. Theoretical objections to this constant, such as its apparent inconsistency with modern particle physics~\cite{Weinberg:1988cp} and seemingly fine-tuned initial conditions~\cite{Zlatev:1998tr}, have long inspired speculation on a more fundamentally sound explanation for the observed late-time expansion~\cite{Joyce:2014kja,Bull:2015stt}. Despite this, no alternative to $\Lambda$ has garnered widespread consensus, and the current cosmological standard model ($\Lambda$CDM) assumes the cosmological constant hypothesis is exactly correct.

Increasingly in the past decade, precision cosmological data have revealed observational tensions within the $\Lambda$CDM paradigm. These notably include the $> 5\sigma$ disagreement between Hubble constant ($H_0$) measurements from the \textit{Planck} cosmic microwave background (CMB) satellite~\cite{Planck:2018vyg} and local distance-ladder estimates from the SH0ES collaboration~\cite{Riess:2021jrx,Murakami:2023xuy,Breuval:2024lsv}, as well as the tension between the galactic-scale matter clustering inferred by \textit{Planck}~\cite{Planck:2018vyg} (quantified by the $S_8$ parameter~\cite{DiValentino:2020vvd,DiValentino:2018gcu}) 
and estimates derived directly from galaxy surveys and weak gravitational lensing observations~\cite{DES:2021bvc,DES:2021vln,KiDS:2020suj,Asgari:2019fkq,Joudaki:2019pmv,DAmico:2019fhj,Kilo-DegreeSurvey:2023gfr,Troster:2019ean,Heymans:2020gsg,Dalal:2023olq,Chen:2024vvk,ACT:2024okh,Sailer:2024coh,DES:2024oud,Harnois-Deraps:2024ucb,Dvornik:2022xap,DES:2021wwk}, which was recently alleviated by the new KiDS-Legacy release~\cite{Wright:2025xka}.
Nonetheless, such inconsistencies suggest that the assumptions of the standard cosmological model could be refined or replaced with ones that have greater observational support.\footnote{See~\cite{DiValentino:2021izs,Abdalla:2022yfr,Perivolaropoulos:2021jda,Khalife:2023qbu,DiValentino:2024yew} for reviews of attempts in this direction.} 

When investigating alternatives to $\Lambda$, two phenomenological approaches are commonly employed. The first is \textit{parametric} and involves assuming a specific functional form for, e.g., the dark energy equation of state, $w(a)$, as a function of cosmic time, here measured by the scale factor $a$. This function typically includes one or more free parameters, which are constrained using observational data~\cite{Cooray:1999da,Efstathiou:1999tm,Astier:2000as,Weller:2001gf,Corasaniti:2002vg,Wetterich:2004pv,Bassett:2004wz,Feng:2004ff,Hannestad:2004cb,Xia:2004rw,Alam:2004jy,Upadhye:2004hh,Gong:2005de,Jassal:2005qc,Nesseris:2005ur,Ananda:2005xp,Linder:2005dw,Nojiri:2006ww,Jain:2007fa,Kurek:2007bu,Ichikawa:2007je,Liu:2008vy,Barboza:2008rh,Barboza:2009ks,Kurek:2008qt,Ma:2011nc,Mukhopadhyay:2007ed,Sendra:2011pt,Feng:2011zzo,Pace:2011kb,Barboza:2011gd,DeFelice:2012vd,Feng:2012gf,Wei:2013jya,Sello:2013oja,Magana:2014voa,Akarsu:2015yea,Pan:2016jli,DiValentino:2016hlg,Nunes:2016plz,Nunes:2016drj,Magana:2017usz,Yang:2017alx,Pan:2017zoh,Panotopoulos:2018sso,Yang:2018qmz,Jaime:2018ftn,Das:2017gjj,Yang:2018prh,Du:2018tia,Li:2019yem,Yang:2019jwn,Pan:2019hac,Tamayo:2019gqj,Rezaei:2019roe,Pan:2019brc,Rezaei:2020mrj,DiValentino:2020naf,Perkovic:2020mph,Banihashemi:2020wtb,Jaber-Bravo:2019nrk,Benaoum:2020qsi,Yang:2021eud,Jaber:2021hho,Alestas:2021luu,Yang:2022klj,Escudero:2022rbq,Castillo-Santos:2022yoi,Yang:2022kho,Dahmani:2023bsb,Escamilla:2023oce,Rezaei:2023xkj,Adil:2023exv,LozanoTorres:2024tnt,Singh:2023ryd,Rezaei:2024vtg,Reyhani:2024cnr,Paliathanasis:2025cuc,Paliathanasis:2025dcr}. Alternatively, \textit{non-parametric} methods use numerical and statistical tools to \textit{reconstruct}, e.g., $w(a)$ in different scale factor or redshift ranges~\cite{Huterer:2002hy,Shafieloo:2005nd,Bogdanos:2009ib,Holsclaw:2010nb,Seikel:2012uu,Nesseris:2012tt,Yahya:2013xma,Zhao:2017cud,Wang:2018fng,Zhang:2019jsu,Escamilla:2021uoj,Escamilla:2024fzq,Jiang:2024xnu,Ye:2024ywg,Ormondroyd:2025exu,Ormondroyd:2025iaf,Berti:2025phi,Gu:2025xie}.
Whereas non-parametric methods are more flexible and involve fewer underlying assumptions~\cite{Huterer:2002hy,Nesseris:2025lke}, parametric models can often achieve tighter constraints on their (fewer) degrees of freedom. Within the parametric approach, there are two standard choices for the form of the dark energy equation of state: the constant $w$CDM model and the dynamical $w_0w_a$ or Chevallier–Polarski–Linder (CPL) parametrization~\cite{Linder:2002et,Chevallier:2000qy}, which arise from Taylor expanding a general $w(a)$ to zeroth and linear order, respectively.\footnote{The CPL parametrization was also proposed to capture the behavior of more fundamental (scalar field) models~\cite{Linder:2002et}.}

Assuming the CPL parametrization, the Dark Energy Spectroscopic Instrument (DESI) collaboration
recently reported a $2.8$–$4.2\sigma$ preference for dynamical dark energy
when combining their own baryon acoustic oscillation (BAO) measurements~\cite{DESI:2025zgx,Lodha:2025qbg,Andrade:2025xyg} with CMB data from \textit{Planck}~\cite{Planck:2018vyg} and different Type Ia supernovae (SNe) catalogs~\cite{Scolnic:2021amr,Brout:2022vxf,DES:2024hip,DES:2024jxu,DES:2024upw,Rubin:2023ovl}. 
The first DESI data release~\cite{DESI:2024mwx,DESI:2024kob,DESI:2024aqx} prompted much discussion on the robustness and implications of these BAO and SNe data~\cite{Cortes:2024lgw,Shlivko:2024llw,Luongo:2024fww,Yin:2024hba,Gialamas:2024lyw,Dinda:2024kjf,Wang:2024dka,Tada:2024znt,Carloni:2024zpl,Chan-GyungPark:2024mlx,Ramadan:2024kmn,Notari:2024rti,Orchard:2024bve,Hernandez-Almada:2024ost,Pourojaghi:2024tmw,Reboucas:2024smm,Giare:2024ocw,Chan-GyungPark:2024brx,Menci:2024hop,Li:2024qus,Li:2024hrv,Notari:2024zmi,Gao:2024ily,Fikri:2024klc,Jiang:2024xnu,Zheng:2024qzi,Gomez-Valent:2024ejh,RoyChoudhury:2024wri,Lewis:2024cqj,Bhattacharya:2024hep, Bhattacharya:2024kxp,Wolf:2024eph,Wolf:2024stt,Shajib:2025tpd}.
Importantly,  
the evidence for dynamical dark energy was confirmed using dataset combinations that do not include DESI BAO measurements~\cite{Giare:2025pzu}\footnote{Formerly, the strength of evidence was driven largely by the SNe catalog. However, DESI-DR2 BAO and CMB data now present $\sim3\sigma$ evidence for dynamical dark energy~\cite{DESI:2025zgx, Lodha:2025qbg}.} and parameterizations other than CPL~\cite{Najafi:2024qzm,Giare:2024gpk,Wolf:2025jlc}.

There are several reasons to consider parameterizations beyond CPL when evaluating the evidence for a dynamical $w(a)$. While this parametrization is limited to a purely linear evolution and diverges in the infinite future ($a \rightarrow \infty$), alternatives such as the Barboza–Alcaniz proposal~\cite{Barboza:2008rh} address both of these potential shortcomings. Notably, this alternative was favored over CPL by the datasets used in the original DESI analysis~\cite{Giare:2024gpk}.
Furthermore, non-parametric reconstructions of the dark energy equation of state consistently find oscillating features during late times ($a \gtrsim 2/3$)~\cite{Zhao:2017cud,Wang:2018fng,Zhang:2019jsu,Escamilla:2021uoj,Escamilla:2024fzq,Jiang:2024xnu,Ye:2024ywg,Ormondroyd:2025exu,Ormondroyd:2025iaf,Berti:2025phi,Gu:2025xie}, this feature being inconsistent with linear evolution and in qualitative agreement with the general parametrization in~\cite{Jaber:2021hho}. In view of these preferences, we consider alternative parametrizations for $w(a)$ that evolve non-monotonically with $a$ and are well-defined throughout all cosmic history. 

An important consideration in phenomenological studies of dark energy is the number of free parameters introduced in the equation of state. This decision is less ambiguous in more fundamental approaches, where parameters come from the underlying microphysics~\cite{Copeland:2006wr}, and in the $w$CDM and CPL parametrizations, which follow from a general series expansion of $w(a)$. For alternative phenomenological models, the optimal number of parameters depends on the goals of the study. When attempting to constrain multiple independent features of $w(a)$ (e.g., its present-day value, phantom crossing, and oscillations), having two or more parameters is justified and likely required for accurate conclusions.
However, models with more phenomenological parameters often have a more degenerate parameter space and worse Bayesian evidence, as the necessarily agnostic priors on the new parameters significantly increase the prior volume. Because our concern is identifying alternatives to $\Lambda$ that provide a better fit to modern cosmological data according to both the chi-squared statistic and Bayesian evidence, we consider parameterizations of the dark energy equation of state with only a single free parameter. Our results demonstrate that some such models are preferred over traditional parametrizations by current CMB, BAO, and Type Ia SNe data.

This paper is organized as follows. In Section~\ref{sec:dark-energy}, we review the cosmological effects of dark energy and introduce the models considered in this work. Section~\ref{sec:data} presents the observational data used in our analyses, and Section~\ref{sec:results} discusses the resulting observational constraints on the dark energy parametrizations. Finally, Section~\ref{sec:summary} summarizes our main findings and concludes.

\section{Dark Energy}
\label{sec:dark-energy}

\begin{figure*}
    \centering
    \includegraphics[width=\linewidth]{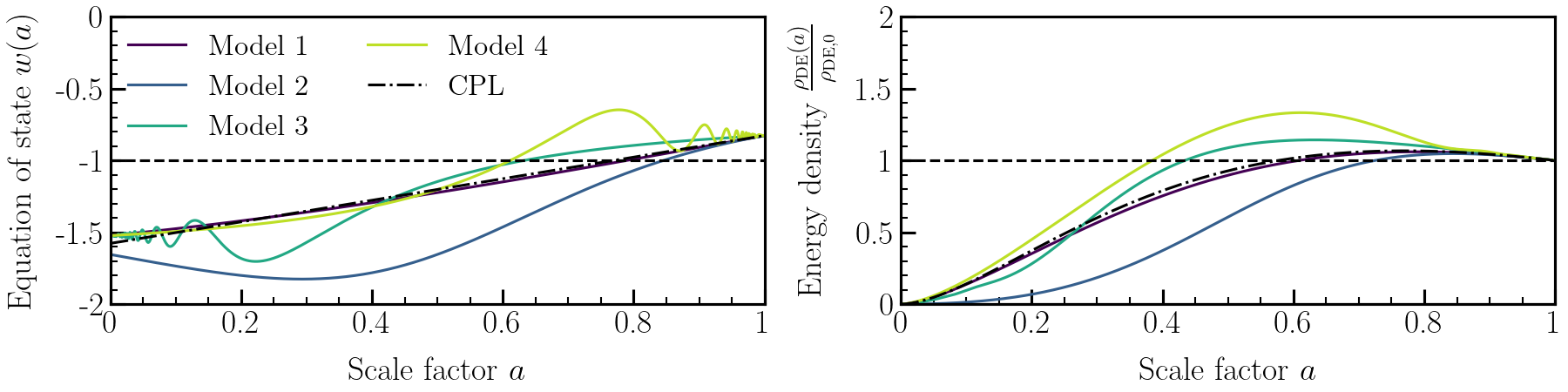}
    \caption{The $w(a)$ and $\rho_{\rm DE}(a)/\rho_{\rm DE,0}$ for each dark energy model considered in this work [Eqs.~\eqref{eq:model1}--\eqref{eq:model4}] are compared with the mean quantities obtained by DESI~\cite{DESI:2025zgx} using the CPL parametrization and the \textit{Planck} 2018, ACT-DR6 Lensing, DESI-DR2 BAO, and PantheonPlus dataset combination (dash-dot lines). Each dark energy model is shown with the same present-day value of $w(a)$, and the horizontal dashed lines represent the $\Lambda$CDM predictions. }
\label{fig:models}
\end{figure*}

Assuming a spatially flat Friedmann-Lema\^itre-Robertson-Walker (FLRW) universe, general relativity predicts the Friedmann equation
\begin{align}
    H(a) = H_0\bigg[\Omega_m a^{-3} + \Omega_\gamma a^{-4}
    + \Omega_\nu \frac{\rho_\nu(a)}{\rho_{\nu,0}} + \Omega_{\rm DE} \frac{\rho_{\rm DE}(a)}{\rho_{\rm DE,0}}\bigg]^{1/2}
    \label{eq:friedman}
\end{align}
for the Hubble parameter $H(a) \equiv (d a / dt)/a$, where $t$ is cosmic time, $a$ is the scale factor, $\Omega_{\cdot} \equiv \rho_{\cdot,0}/(3H_0^2/8\pi G)$, $\Omega_m \equiv \Omega_b + \Omega_c$, and we have chosen the present-day scale factor $a_0 \equiv 1$. In these equations, a zero subscript implies evaluation at present, while the subscripts $b,\,c,\,\gamma,\,\nu,$ and DE refer to baryons, cold dark matter, photons, neutrinos, and dark energy, respectively. Assuming that dark energy does not interact with the other components and is minimally coupled to gravity, its energy density obeys the evolution equation
\begin{equation}
 \frac{\rho_\mathrm{DE}(a)}{\rho_{\mathrm{DE},0}} = a^{-3} \exp\left[3 \int_a^{1} da'\: \frac{w(a')}{a'}\right]\,,
 \label{eq:DE-evolution}
\end{equation}
where $w(a)$ is the (barotropic) dark energy equation of state. Given a specific model for $w(a)$, Eqs.~\eqref{eq:friedman} and \eqref{eq:DE-evolution} determine the background geometry of the universe.

To model dark energy perturbations, we use the default parameterized post-Friedmann (PPF) method implemented in the Boltzmann code \texttt{CAMB}~\cite{Lewis:1999bs}, which allows $w(a)$ to cross the phantom divide ($w = -1$) without introducing divergences in the perturbation equations~\cite{Hu:2007pj,Fang:2008sn}. 
In the PPF framework, a free function $\Gamma$ parametrizes deviations from the evolution of metric perturbations in a universe without dark energy. By requiring that these deviations are consistent with a spatially flat FLRW background on super-horizon scales, satisfy local energy–momentum conservation, and are fully suppressed on small scales, Refs.~\cite{Hu:2007pj,Fang:2008sn} derive the defining and evolution equations for $\Gamma$:

\begin{align}
    &\Gamma = -4\pi G \left(\frac{a}{k}\right)^2 \rho_{\rm DE}\, \delta_{\rm DE}^\mathrm{(rest)}, \\
    &(1 + c_\Gamma^2 k_H^2)\, \left[\frac{d\,\Gamma}{d(\ln a)} + (1 + c_\Gamma^2 k_H^2)\, \Gamma\right] = S,
    \label{eq:Gamma-evolution}
\end{align}
where $\delta_{\rm DE}^\mathrm{(rest)} \equiv (\delta \rho_{\rm DE} / \rho_{\rm DE})^\mathrm{(rest)}$ is evaluated in the dark energy rest frame, and $c_\Gamma$ determines the scale at which dark energy becomes smooth relative to matter.\footnote{Calibrating the PPF formalism on scalar field models of dark energy gives $c_\Gamma = 0.4\, c_\mathrm{DE}$~\cite{Fang:2008sn}, where $c_\mathrm{DE}$ is the dark energy sound speed. This is the default value chosen by \texttt{CAMB}.} The quantity $k_H = k / (aH)$ is the physical wavenumber relative to the Hubble parameter, and the source term $S$ in Eq.~\eqref{eq:Gamma-evolution} includes contributions from $\rho_{\rm DE}(a)$ and $w(a)$, alongside velocity perturbations in the matter sector~\cite{Fang:2008sn}.

\subsection{Models}

There are, necessarily, many more potential one-parameter models of $w(a)$ in the literature\footnote{These include models that were one-parameter at their conception~\cite{Gong:2005de,Yang:2018qmz,Yang:2021eud} as well as models that can be reduced to a single parameter in the manner described later in this section.} 
than we can reasonably analyze here. For this reason, we restrict our analysis to a representative subset of four models whose selection was guided by the following principles: the parametrization (\textit{i}) is simple, involving only elementary functions and a limited number of factors, (\textit{ii}) contains departures from the CPL form with a phenomenological interpretation, and (\textit{iii}) has demonstrated promise in previous statistical analyses. For ease of comparison between the models and CPL, we further required that each parametrization be of the form
\begin{equation}
    w(a) = w_0\, g(a)\,, \qquad g(1)=1\,,
\end{equation}
so that the single parameter $w_0$ always corresponds to the present-day value of the equation of state. Importantly, our aim is not to perform an unbiased Bayesian comparison between dark energy models with one parameter versus those with more or fewer, where the choice of models could introduce bias. Rather, we seek to identify phenomenologically sound one-parameter cases that can provide insight into the nature of dark energy. Achieving a statistical improvement over standard dark energy models ($\Lambda$CDM and CPL) is straightforward with abundant parameters, where overfitting becomes possible, but non-trivial in a single-parameter framework. That some of the parametrizations considered here do yield improvements is therefore instructive for the construction of future theoretical models.  
As we elaborate below, principle (\textit{ii}) naturally leads us to parametrizations that are well defined across cosmic history and exhibit departures from linear evolution, as often suggested by non-parametric reconstructions~\cite{Zhao:2017cud,Wang:2018fng,Zhang:2019jsu,Escamilla:2021uoj,Escamilla:2024fzq,Jiang:2024xnu,Ye:2024ywg,Ormondroyd:2025exu,Ormondroyd:2025iaf,Berti:2025phi,Gu:2025xie}, while (\textit{iii}) leads us to the parametric shapes proposed by Barboza--Alcaniz~\cite{Barboza:2008rh} and Ma--Zhang~\cite{Ma:2011nc}.

We consider four one-parameter models of the dark energy equation of state:
\begin{align}
    \text{Model 1:}\quad & w(a) = w_0\big[1 + \sin(1 - a)\big] \label{eq:model1} \\
    \text{Model 2:}\quad & w(a) = w_0 \left[1 +  \frac{1 - a}{a^2 + (1 - a)^2} \right]  \label{eq:model2} \\
    \text{Model 3:}\quad & w(a) = w_0 \left[1 - a \sin \left(\frac{1}{a} \right) + \sin 1 \right]  \label{eq:model3} \\
    \text{Model 4:}\quad & w(a) = w_0 \left[1 + (1 - a) \sin\left(\frac{1}{1 - a} \right) \right]. \label{eq:model4}
\end{align}

\begin{figure*}
    \begin{tabular}{c l}
        (a) & \includegraphics[width=\linewidth]{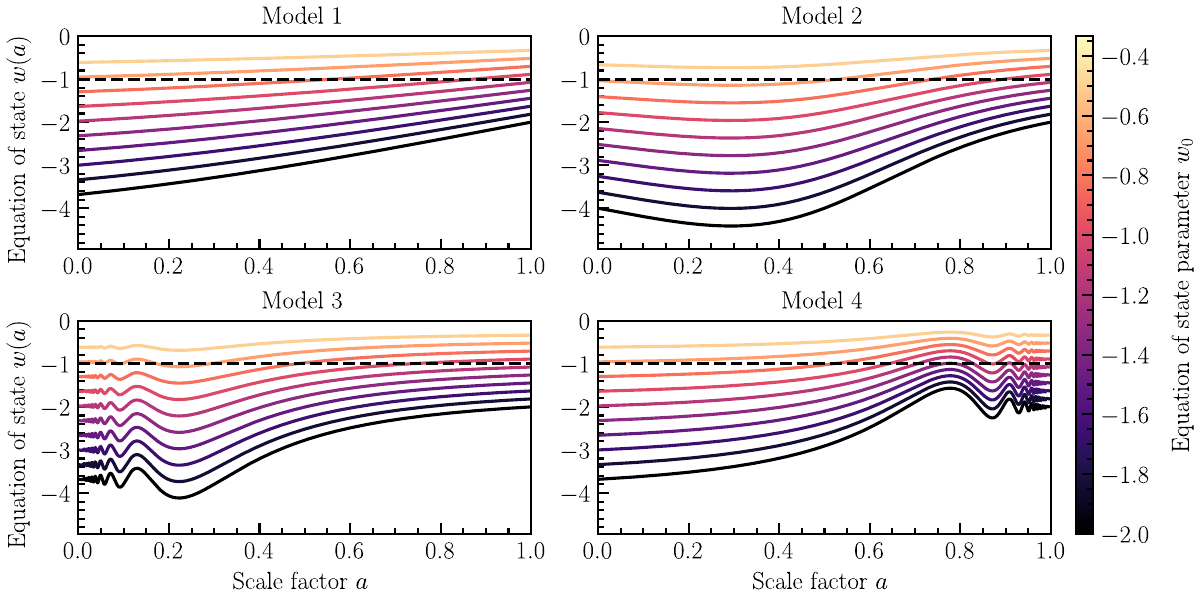}\label{fig:EoS-gradient} \\
        (b) & \includegraphics[width=0.9\linewidth]{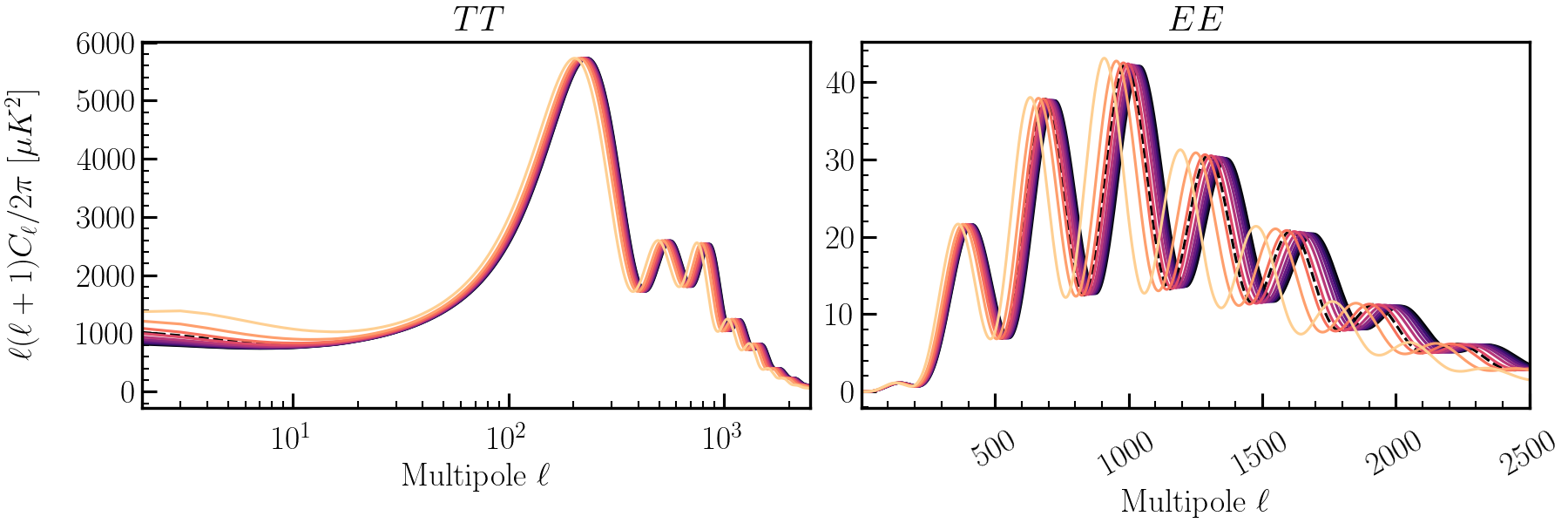}\label{fig:ps-gradient}
    \end{tabular}  
  \caption{(a) The equation of state of each dark energy model when $w_0$ is varied in the range $[-2, -1/3]$. (b) The corresponding effect of Model~\hyperref[eq:model1]{1} on the CMB TT and EE power spectra. Here, $w_0$ is varied while fixing the six $\Lambda$CDM parameters to their \textit{Planck} 2018 values, and the dashed lines show the best-fitting $\Lambda$CDM spectra. The results for the other three models are similar, with small differences ($\lesssim 5\%$) in the low-$\ell$ TT spectrum, due to the ISW effect, and slightly larger differences ($\lesssim 10\%$) in the high-$\ell$ TT spectrum, due to the geometrical degeneracy between $w_0$ and $H_0$, when $w_0$ is restricted to the range $w_0 \in [-1, -0.65]$ that is preferred by our analysis (Section~\ref{sec:results}).}\label{fig:gradient}
\end{figure*}

The equation of state and the corresponding energy density, $\rho_{\rm DE}(a)$, for each model are depicted in Fig.~\ref{fig:models}. These models have increasing shapes\footnote{Decreasing equations of state are not considered, as they are disfavored by current cosmological data: using the CPL parametrization, the DESI collaboration found that $w_a < 0$ at greater than $2.5\sigma$ significance. We have further confirmed that the decreasing versions of Models 1 and 4 provide significantly worse fits to the datasets we consider.} that can be shifted vertically and modulated in amplitude by varying their single parameter, $w_0$, which also represents the present-day value of $w(a)$. The variation of each equation of state with $w_0$ is shown in the top panel of Fig.~\ref{fig:gradient}. Although every model is well-defined and includes some inflection points over the entire history of the universe, corresponding to $a \in [0,\,\infty)$, only Models 2 through 4 have 
inflections within the time domain $a \in [0, 1]$ that is relevant to our analysis. 

 A similar version of the first parametrization (Model~\hyperref[eq:model1]{1}) was explored in~\cite{Yang:2018qmz}  and it arises from a simple elementary function (the sine function) that does not require an additional parameter to control the slope: its average slope naturally aligns with the CPL preference found by DESI~\cite{DESI:2025zgx}. This model has the lowest frequency of oscillations and is the only one to remain monotonic over $a \in [0, 1]$. Model~\hyperref[eq:model2]{2} can be viewed as a restricted version of the parametric shape proposed by Barboza–Alcaniz ~\cite{Barboza:2008rh}, which defines a $g(a)$ that is similar to the CPL form but remains finite over all cosmic times.
This model has a single inflection point, decreasing until $a \sim 0.3$ before increasing toward the present day.

The last two equations of state (Models~\hyperref[eq:model3]{3} and~\hyperref[eq:model4]{4}) are inspired by the two-parameter oscillating parametrization introduced by Ma and Zhang~\cite{Ma:2011nc}. In both models, a linear envelope is supplemented by oscillations that rapidly increase in frequency after the beginning of the universe (Model~\hyperref[eq:model3]{3}) or before the present day (Model~\hyperref[eq:model4]{4}). Whereas the former model is obtained by equating the parameters of the Ma–Zhang equation of state, the latter has (to our knowledge) not been analyzed previously in the literature. Model~\hyperref[eq:model4]{4} is further notable for its ability to capture both the phantom crossing preferred by parametric studies (e.g., ones by DESI~\cite{DESI:2024mwx,DESI:2024aqx,DESI:2025zgx,Lodha:2025qbg}) and the late-time extrema and oscillations found in non-parametric reconstructions~\cite{Zhao:2017cud,Wang:2018fng,Zhang:2019jsu,Escamilla:2021uoj,Escamilla:2024fzq,Jiang:2024xnu,Ye:2024ywg,Ormondroyd:2025exu,Ormondroyd:2025iaf,Berti:2025phi,Gu:2025xie}.

It is worth restating that Models 2 and 3 were not single-parameter
models in their original formulations, but rather had two free
parameters. Previous observational analyses found degeneracy between these parameters, motivating the introduction of a one-parameter form. These simplifications thus come more from previous data than \textit{a priori} physical considerations.

The bottom panel of Fig.~\ref{fig:gradient} illustrates the impact of these dark energy models on the CMB TT and EE power spectra. Explicitly, these spectra were obtained assuming Model~\hyperref[eq:model1]{1}. However, the other models yield similar results, with the most noticeable differences appearing in the low-$\ell$ plateau (due to the late-time Integrated Sachs–Wolfe effect) and the high-$\ell$ damping tail (due to the geometrical degeneracy between $w_0$ and $H_0$). These differences are limited to around $5\%$ and $10\%$, respectively, when the equation of state parameter is restricted to the range $w_0 \in [-1, -0.65]$ that is preferred by our analysis (Section~\ref{sec:results}). Since Model~\hyperref[eq:model3]{3} affects the CMB in much the same way as Model~\hyperref[eq:model1]{1}, its early-time oscillations are not expected to significantly affect its fit to the datasets we consider. Conversely, the late-time oscillations of Model~\hyperref[eq:model4]{4} should be resolvable through the BAO and SNe distance measurements.

\section{Observational Data and Methodology}
\label{sec:data}

To perform parameter inference, we use the publicly available Markov Chain Monte Carlo (MCMC) sampler \texttt{Cobaya}~\cite{Torrado:2020dgo} in conjunction with the Boltzmann solver \texttt{CAMB}~\cite{Lewis:1999bs}, modified to incorporate our dark energy parameterizations. We assess the convergence of our MCMC chains using the Gelman–Rubin diagnostic parameter, $R - 1$~\cite{Gelman:1992zz}, and consider the chains to be converged when the criterion $R - 1 < 0.02$ is met. The MCMC results are analyzed and plotted using \texttt{getdist}~\cite{Lewis:2019xzd}.

Our models extend the standard $\Lambda$CDM framework by introducing an additional parameter for the dark energy equation of state, bringing the total number of free parameters to seven. These are: the physical baryon density $\Omega_b h^2$, the physical dark matter density $\Omega_c h^2$, the optical depth to reionization $\tau$, the angular size of the sound horizon at recombination $\theta_s$, the amplitude of primordial scalar perturbations $\log{(10^{10} A_s)}$, the scalar spectral index $n_s$, and the equation of state parameter $w_0$. 
For all parameters, we assume the flat (uninformative) priors given in Table~\ref{table:priors}.

\begin{table}
	\begin{center}
		\footnotesize
		\renewcommand{\arraystretch}{1.5}
		\begin{tabular}{l@{\hspace{0. cm}}@{\hspace{2 cm}} c}
			\hline\hline
			\textbf{Parameter} & \textbf{Prior} \\
			\hline\hline
			$\Omega_{\rm b} h^2$ & $[0.005\,,\,0.1]$ \\
			$\Omega_{\rm c} h^2$ & $[0.005\,,\,0.99]$ \\
			$\tau$ & $[0.01, 0.8]$ \\
			$100\,\theta_s$ & $[0.5\,,\,10]$ \\
			$\log(10^{10}A_{\rm S})$ & $[1.61\,,\,3.91]$ \\
			$n_{\rm s}$ & $[0.8\,,\, 1.2]$ \\
			$w_{0}$ & $[-2\,,\,0]$ \\
			\hline\hline
		\end{tabular}
		\caption{The flat prior distributions imposed on the cosmological parameters in our analyses. We assume the prior $w_a \in [-3, 2]$ for the additional parameter of the CPL model.}
		\label{table:priors}
	\end{center}
\end{table}

To constrain these parameters, we use the following datasets:

\begin{itemize}
    \item Cosmic Microwave Background (CMB) temperature anisotropy and polarization power spectra, their cross-spectra, and the reconstructed lensing from the \textit{Planck} 2018 legacy data release~\cite{Planck:2018lbu,Planck:2018nkj,Planck:2018vyg,Planck:2019nip}. This dataset is referred to as \textbf{\textit{Planck} 2018}.
    
    \item The sixth data release of the CMB lensing power spectrum from the Atacama Cosmology Telescope~\cite{ACT:2023kun,ACT:2023dou}, which incorporates measurements from \textit{Planck}.\footnote{We use the \texttt{actplanck\_baseline} likelihood variant from \href{https://github.com/ACTCollaboration/act\_dr6\_lenslike}{https://github.com/ACTCollaboration/act\_dr6\_lenslike}.} This dataset is referred to as \textbf{ACT-DR6 Lensing}.
    
    \item Baryon Acoustic Oscillation (BAO) measurements from the first two years of observations by the Dark Energy Spectroscopic Instrument (DESI)~\cite{DESI:2025zgx,Lodha:2025qbg,Andrade:2025xyg}. This dataset is referred to as \textbf{DESI-DR2 BAO}.
\end{itemize}

These three datasets (\textit{Planck} 2018, ACT-DR6 Lensing, and DESI-DR2 BAO) form our “baseline” for parameter inference and are included in all analyses. In contrast, the next three datasets consist of different Type Ia SNe samples, which are not used simultaneously.
These are:

\begin{itemize}
    \item A total of 1701 light curves from 1550 distinct SNe spanning the redshift range $0.01 < z < 2.26$, obtained from the PantheonPlus sample~\cite{Scolnic:2021amr,Brout:2022vxf}. This dataset is referred to as \textbf{PantheonPlus}.
    
    \item The full 5-year dataset of the Dark Energy Survey (DES) Supernova Program, which includes distance modulus measurements for 1829 SNe in the redshift range $0.02 < z < 1.13$~\cite{DES:2024hip,DES:2024jxu,DES:2024upw}. This dataset is referred to as \textbf{DESY5}.\footnote{After the completion of this paper, several systematics in DESY5 were identified and corrected in the updated DES-Dovekie SNe sample~\cite{Popovic:2025glk, DES:2025sig}. See the \textit{Added note} following Sec.~\ref{sec:summary} for a discussion of how these changes affect our results.} 
    \item The Union3 compilation, consisting of 2087 SNe~\cite{Rubin:2023ovl}. This dataset is referred to as \textbf{Union3}.
\end{itemize}

\subsection{\label{sec:preference-stats}Model Preference Statistics}

\begin{table}
\scalebox{0.75}{ 
\begin{tabular} { l  c c c}
\noalign{\vskip 3pt}\hline\noalign{\vskip 1.5pt}\hline\noalign{\vskip 5pt}
 \multicolumn{1}{c}{\bf Parameter} &  \multicolumn{1}{c}{\bf PantheonPlus} &  \multicolumn{1}{c}{\bf DESY5} &  \multicolumn{1}{c}{\bf Union3}\\\noalign{\vskip 3pt}\hline\noalign{\vskip 1.5pt}\hline\noalign{\vskip 5pt}
{$\Omega_\mathrm{c} h^2$} & $0.11932\pm 0.00075        $ & $0.11898\pm 0.00075        $ & $0.11884\pm 0.00077        $\\

{$\Omega_\mathrm{b} h^2$} & $0.02243\pm 0.00013        $ & $0.02245\pm 0.00013        $ & $0.02247\pm 0.00013        $\\

{$100\theta_\mathrm{MC}$} & $1.04100\pm 0.00028        $ & $1.04105\pm 0.00029        $ & $1.04107\pm 0.00029        $\\

{$\tau_\mathrm{reio}$} & $0.0553\pm 0.0071          $ & $0.0565\pm 0.0071          $ & $0.0570^{+0.0068}_{-0.0077}$\\

{$n_\mathrm{s}   $} & $0.9672\pm 0.0034          $ & $0.9680\pm 0.0035          $ & $0.9683\pm 0.0035          $\\

{$\log(10^{10} A_\mathrm{s})$} & $3.045\pm 0.013            $ & $3.047\pm 0.013            $ & $3.049\pm 0.013            $\\

{$w_0            $} & $-0.795\pm 0.021           $ & $-0.773\pm 0.019           $ & $-0.762\pm 0.024           $\\

{$\Omega_\mathrm{m}$} & $0.3126\pm 0.0053          $ & $0.3175\pm 0.0051          $ & $0.3199\pm 0.0062          $\\

{$\sigma_8       $} & $0.8125\pm 0.0086          $ & $0.8053\pm 0.0081          $ & $0.8021\pm 0.0095          $\\

{$S_8            $} & $0.8293\pm 0.0076          $ & $0.8284\pm 0.0077          $ & $0.8282\pm 0.0077          $\\

{$H_0            $} & $67.50\pm 0.59             $ & $66.90\pm 0.54             $ & $66.62\pm 0.68             $\\

{$r_\mathrm{drag}$} & $147.22\pm 0.20            $ & $147.28\pm 0.20            $ & $147.30\pm 0.21            $\\
\noalign{\vskip 3pt}\hline\noalign{\vskip 3pt}
 $\Delta\chi^2_{\rm min}\:(\Delta\ln B)\:{\Lambda\rm CDM}$ & $-$8.2 ($-$0.0) & $-$19.3 (5.6) & $-$12.8 (3.3) \\ 
 $\Delta\chi^2_{\rm min}\:(\Delta\ln B)\:\mathrm{CPL}$ & 1.0 (1.8) & $-$0.4 (2.1) & 0.8 (1.2) \\ 
 \noalign{\vskip 3pt}\hline\noalign{\vskip 3pt} 
 \end{tabular}
 }

\caption{\textbf{Model 1}. Constraints (68\% CL) on the seven free parameters listed in Table~\ref{table:priors}, alongside relevant derived parameters, using the combination of our baseline dataset with each of the three Type Ia SNe samples (PantheonPlus, DESY5, and Union3) separately. The minimum chi-squared difference, $\Delta\chi^2_{\rm min}$ [Eq.~\eqref{eq:chi-squared}], and the logarithmic Bayesian evidence ratio, $\Delta \ln B$ [Eq.~\eqref{eq:bayes-ratio}], are reported at the end of the table.}
\label{tab:model1}

 \end{table}

  \begin{figure}
    \centering
    \includegraphics[width=\linewidth]{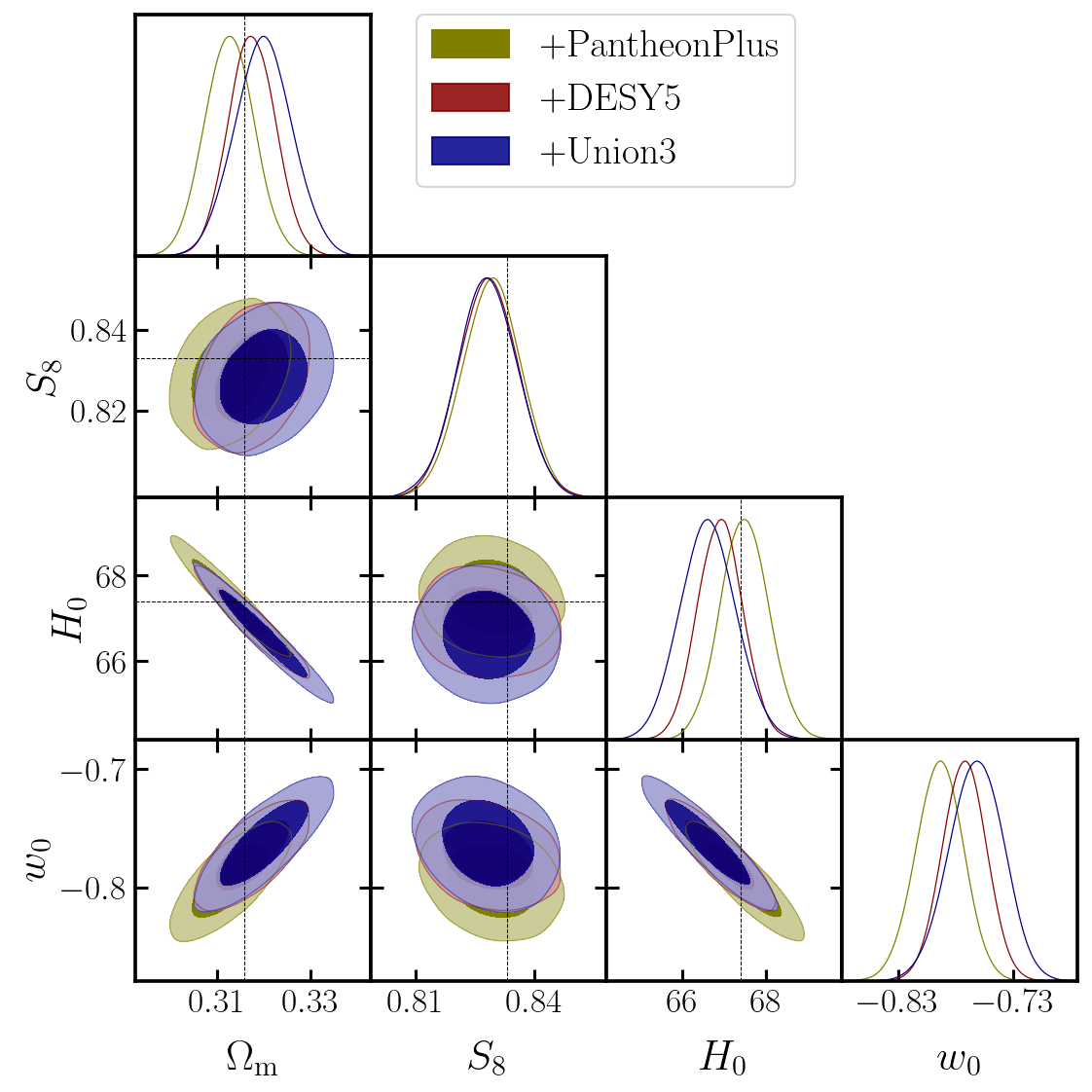}
    \includegraphics[width=\linewidth]{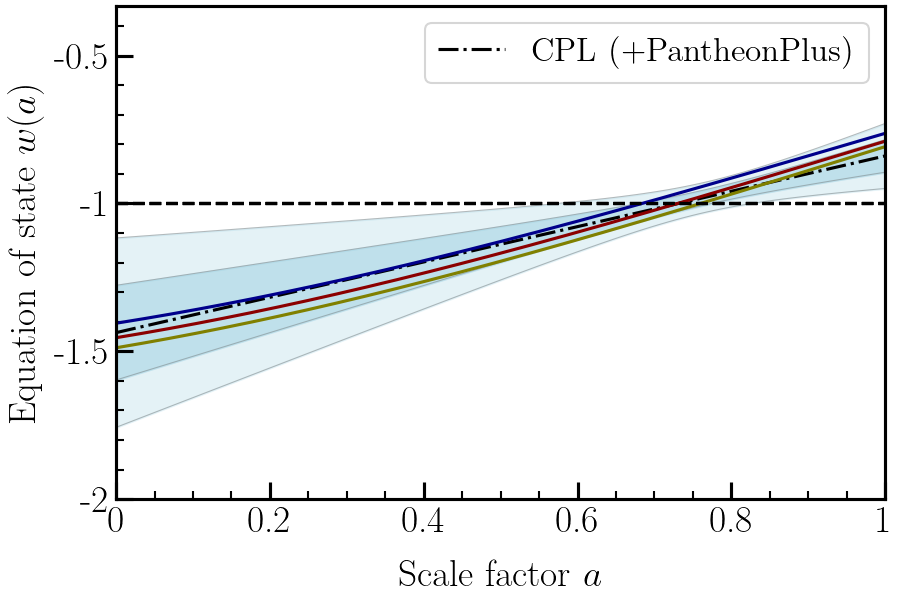}
    \caption{\textbf{Model 1}. \textit{Above:} Posterior distributions and $1\sigma$, $2\sigma$ contours for three key cosmological parameters—the dark energy equation of state parameter $w_0$, the Hubble constant $H_0$, and the matter fluctuation parameter $S_8$—obtained from the combination of our baseline dataset with each of the three Type Ia SNe samples, as indicated in the legend. The dashed lines represent the mean $\Lambda$CDM parameters from \textit{Planck} 2018~\cite{Planck:2018vyg}.  
    \textit{Below:} The best-fitting $w(a)$ for the three dataset combinations shown in the legend above. The dash-dot line shows the mean $w(a)$ obtained using the CPL parametrization with the PantheonPlus combination, while the light blue bands indicate the corresponding $1\sigma$ and $2\sigma$ intervals.}
    \label{fig:model1}
\end{figure}

To quantify the preference (or lack thereof) of these datasets for the dark energy models in Eqs.~\eqref{eq:model1}--\eqref{eq:model4} compared to $\Lambda$CDM and the CPL parametrization, we use two statistical measures: the change in the minimum (best-fitting) chi-squared, $\Delta\chi^2_{\rm min}$, and the logarithmic Bayesian evidence ratio, $\Delta\ln B$:
\begin{align}
    \Delta\chi^2_{\rm min} &\equiv \min \chi^2_{\text{Model }i} - \min \chi^2_{\Lambda\text{CDM or CPL}}\,, \label{eq:chi-squared} \\
    \Delta \ln B &\equiv \ln\left[\frac{B_{\text{Model }i}}{B_{\Lambda\text{CDM or CPL}}}\right]\,, \label{eq:bayes-ratio}
\end{align}
where both $\chi^2_{\rm min}$ and the Bayesian evidence $B$ are obtained directly from our MCMC chains, with the latter computed using \texttt{MCEvidence}~\cite{Heavens:2017afc} and the \texttt{Cobaya} wrapper in the \texttt{wgcosmo} repository~\cite{Giare:wgcosmo}. A model provides a better fit to the data according to the chi-squared statistic if $\Delta\chi^2_{\rm min} < 0$,\footnote{We cannot strictly apply Wilks' theorem to perform a likelihood ratio test—deriving a $p$-value and corresponding “$\sigma$” significance from $\Delta\chi^2_{\rm min}$—because $\Lambda$CDM and CPL are not nested within our one-parameter models.} while a Bayesian preference is theoretically indicated by $\Delta\ln B > 0$.

Unlike the minimum chi-squared, the Bayes ratio accounts for the number of free parameters and prior ranges of the models, penalizing those with larger prior volumes and greater complexity. To interpret the Bayesian evidence, we refer to the revised Jeffreys' scale~\cite{Kass:1995loi}, where $|\Delta\ln B| \lesssim 1$ is considered inconclusive, $1 \lesssim |\Delta\ln B| \lesssim 3$ indicates moderate evidence, and $3 \lesssim |\Delta\ln B| \lesssim 5$ corresponds to strong evidence.

\section{Results}
\label{sec:results}

\begin{figure}
    \centering
    \includegraphics[width=\linewidth]{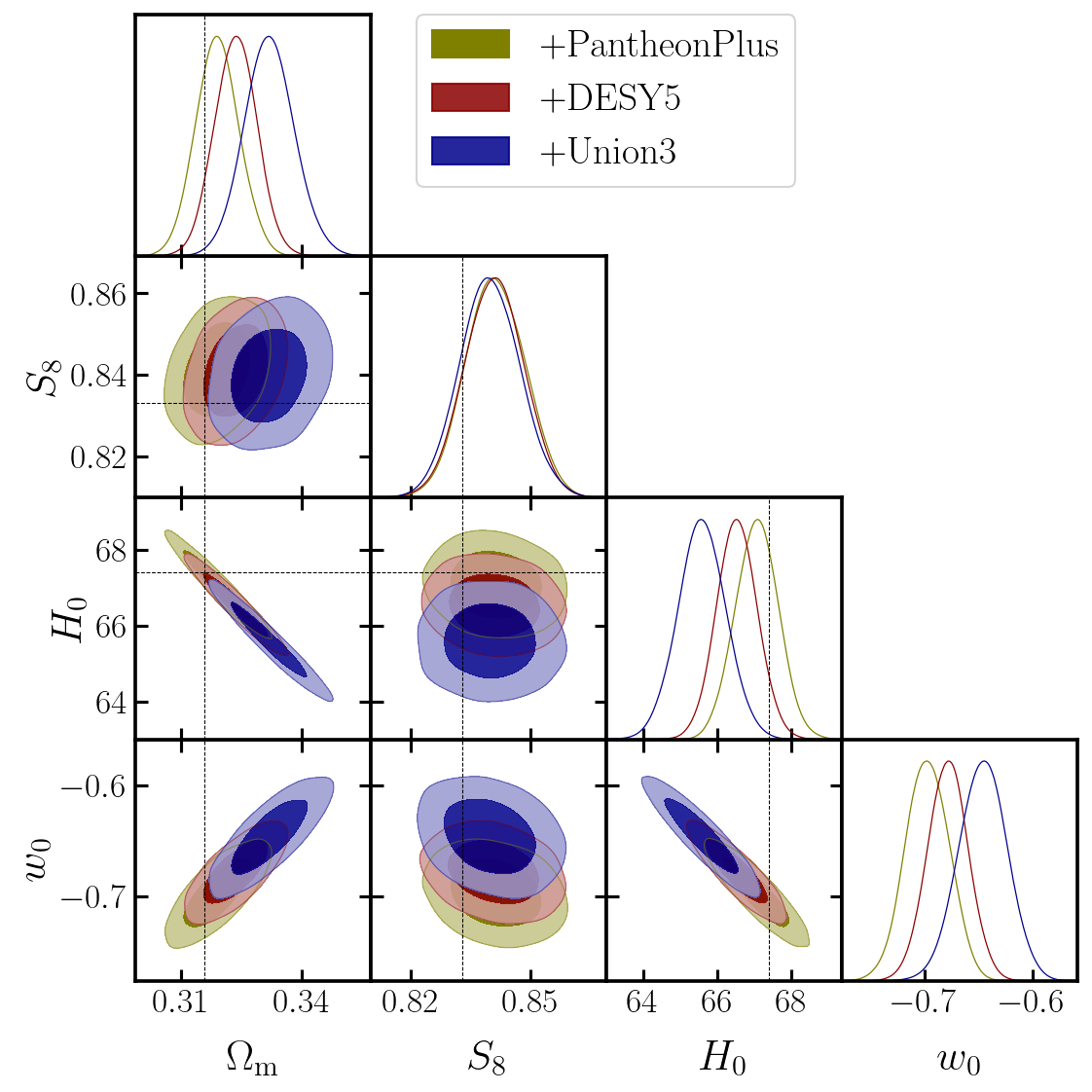}
    \includegraphics[width=\linewidth]{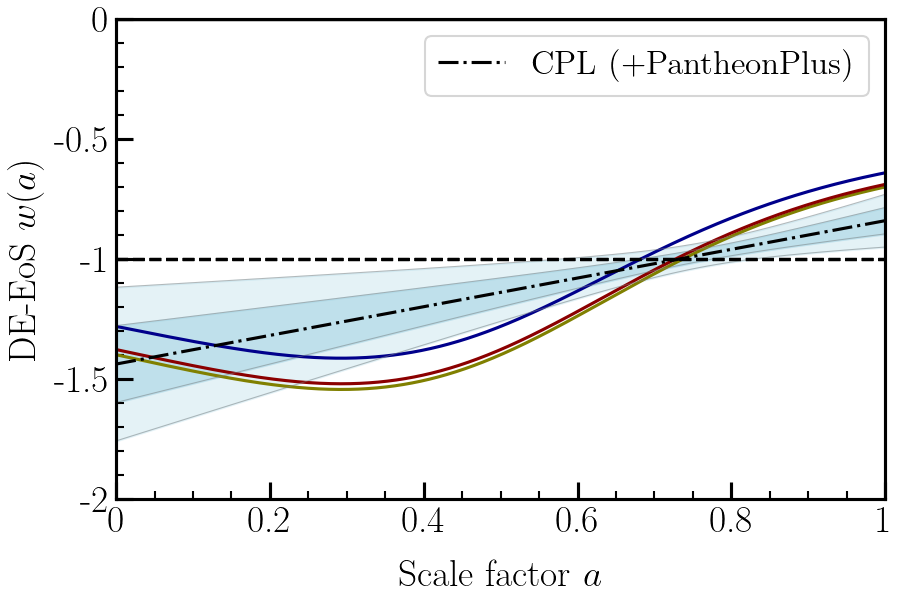}
    
    \caption{\textbf{Model 2}. \textit{Above:} Posterior distributions and $1\sigma$, $2\sigma$ contours for three key cosmological parameters—the dark energy equation of state parameter $w_0$, the Hubble constant $H_0$, and the matter fluctuation parameter $S_8$—obtained from the combination of our baseline dataset with each of the three Type Ia SNe samples, separately, as indicated in the legend. The dashed lines represent the mean $\Lambda$CDM parameter values from \textit{Planck} 2018~\cite{Planck:2018vyg}.  
    \textit{Below:} The best-fitting $w(a)$ for the three dataset combinations shown in the legend above. The dash-dot line shows the mean $w(a)$ obtained using the CPL parametrization with the PantheonPlus combination, while the light blue bands represent the corresponding $1\sigma$ and $2\sigma$ intervals.}
    \label{fig:model2}
\end{figure}

\begin{figure}
    \centering
    \includegraphics[width=\linewidth]{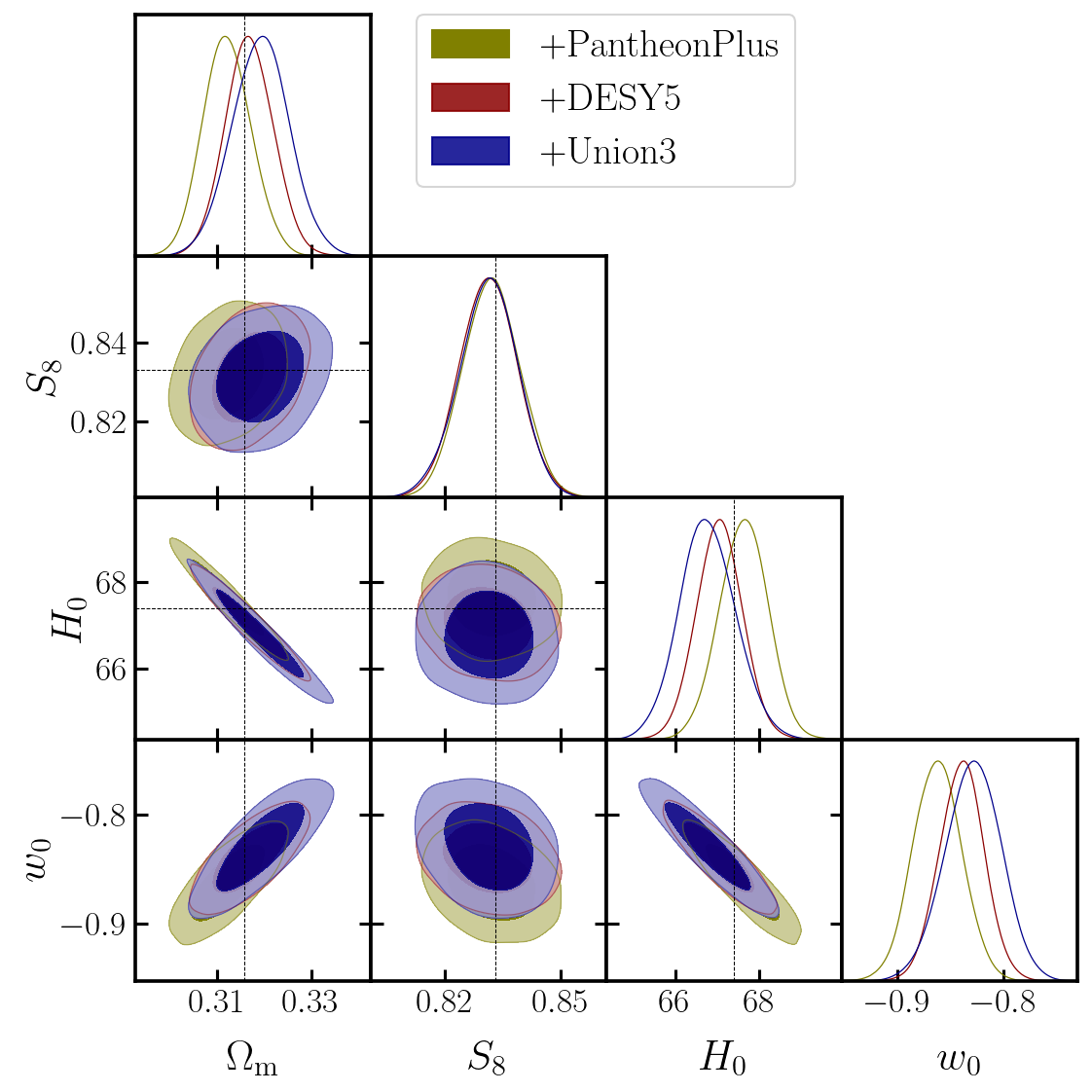}
    \includegraphics[width=\linewidth]{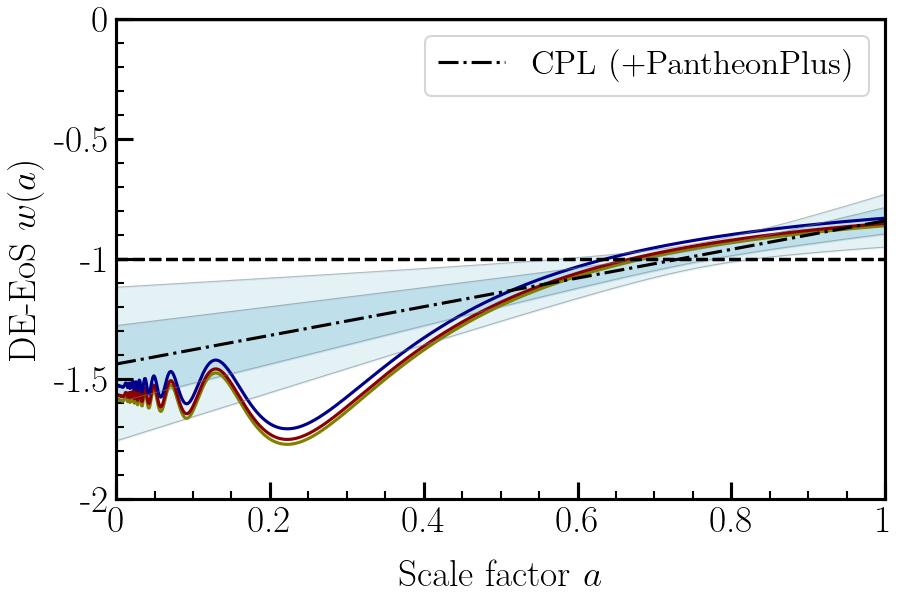}
    
   \caption{\textbf{Model 3}. \textit{Above:} Posterior distributions and $1\sigma$, $2\sigma$ contours for three key cosmological parameters—the dark energy equation of state parameter $w_0$, the Hubble constant $H_0$, and the matter fluctuation parameter $S_8$—obtained from the combination of our baseline dataset with each of the three Type Ia SNe samples, separately, as indicated in the legend. The dashed lines represent the mean $\Lambda$CDM parameters from \textit{Planck} 2018~\cite{Planck:2018vyg}.  
   \textit{Below:} The best-fitting $w(a)$ for the three dataset combinations shown in the legend above. The dash-dot line shows the mean $w(a)$ obtained using the CPL parametrization with the PantheonPlus combination, while the light blue bands represent the corresponding $1\sigma$ and $2\sigma$ intervals.}
    \label{fig:model3}
\end{figure}

In this section, we present observational constraints on the four dark energy models in Eqs.~\eqref{eq:model1}--\eqref{eq:model4} and evaluate their model preference statistics compared to $\Lambda$CDM and the CPL parametrization, using the methodology outlined in the previous section. The posterior distributions of the dark energy equation of state parameter $w_0$, the Hubble constant $H_0$, and the matter fluctuation parameter $S_8$ are shown in Figs.~\ref{fig:model1}--\ref{fig:model4}, alongside the best-fitting dark energy equation of state $w(a)$, for each dataset combination. Observational constraints on each model's energy density $\rho_{\rm DE}(a)$ are compared in Fig.~\ref{fig:rho-de}. Complete information on the models' seven free parameters and relevant derived parameters is given in Tables~\ref{tab:model1}--\ref{tab:model4}. Finally, the model preference statistics are summarized in Table~\ref{tab:preference-stats} and Fig.~\ref{fig:chi2-lnB}, which provides a useful visual depiction.

For robust and comprehensive results, these constraints were obtained using our baseline dataset combined separately with each of the three Type Ia SNe samples (PantheonPlus, DESY5, and Union3). Thus, in the discussions to follow, the choice of SNe catalog serves to identify the dataset combination being considered.

\begin{table}
\scalebox{0.75}{ 
\begin{tabular} { l  c c c}
\noalign{\vskip 3pt}\hline\noalign{\vskip 1.5pt}\hline\noalign{\vskip 5pt}
 \multicolumn{1}{c}{\bf Parameter} &  \multicolumn{1}{c}{\bf +PantheonPlus} &  \multicolumn{1}{c}{\bf +DESY5} &  \multicolumn{1}{c}{\bf +Union3}\\\noalign{\vskip 3pt}\hline\noalign{\vskip 1.5pt}\hline\noalign{\vskip 5pt}
{$\Omega_\mathrm{c} h^2$} & $0.12056\pm 0.00073        $ & $0.12026\pm 0.00074        $ & $0.11977\pm 0.00080        $\\

{$\Omega_\mathrm{b} h^2$} & $0.02233\pm 0.00013        $ & $0.02235\pm 0.00013        $ & $0.02240\pm 0.00013        $\\

{$100\theta_\mathrm{MC}$} & $1.04085\pm 0.00028        $ & $1.04088\pm 0.00029        $ & $1.04095\pm 0.00029        $\\

{$\tau_\mathrm{reio}$} & $0.0503\pm 0.0070          $ & $0.0517\pm 0.0069          $ & $0.0534\pm 0.0070          $\\

{$n_\mathrm{s}   $} & $0.9640\pm 0.0035          $ & $0.9648\pm 0.0034          $ & $0.9661\pm 0.0036          $\\

{$\log(10^{10} A_\mathrm{s})$} & $3.036\pm 0.013            $ & $3.038\pm 0.013            $ & $3.041\pm 0.013            $\\

{$w_0            $} & $-0.697\pm 0.021           $ & $-0.679\pm 0.019           $ & $-0.646\pm 0.023           $\\

{$\Omega_\mathrm{m}$} & $0.3191\pm 0.0055          $ & $0.3236\pm 0.0054          $ & $0.3320\pm 0.0063          $\\

{$\sigma_8       $} & $0.8158\pm 0.0086          $ & $0.8097\pm 0.0083          $ & $0.7985\pm 0.0094          $\\

{$S_8            $} & $0.8413\pm 0.0076          $ & $0.8409\pm 0.0076          $ & $0.8399\pm 0.0078          $\\

{$H_0            $} & $67.07\pm 0.59             $ & $66.54\pm 0.56             $ & $65.60\pm 0.66             $\\

{$r_\mathrm{drag}$} & $147.00\pm 0.20            $ & $147.06\pm 0.20            $ & $147.14\pm 0.21            $\\
\noalign{\vskip 3pt}\hline\noalign{\vskip 3pt}
 $\Delta\chi^2_{\rm min}\:(\Delta\ln B)\:{\Lambda\rm CDM}$ & 5.1 ($-$6.5) & $-$13.1 (2.3) & $-$13.2 (3.2) \\ 
 $\Delta\chi^2_{\rm min}\:(\Delta\ln B)\:\mathrm{CPL}$ & 14.2 ($-$4.6) & 5.7 ($-$1.3) & 0.4 (1.1) \\ 
 \noalign{\vskip 3pt}\hline\noalign{\vskip 3pt} 
 \end{tabular}
}

\caption{\textbf{Model 2}. Constraints (68\% CL) on the seven free parameters listed in Table~\ref{table:priors}, along with relevant derived parameters, obtained using the combination of our baseline dataset with each of the three Type Ia SNe samples (PantheonPlus, DESY5, and Union3) separately. The minimum chi-squared difference, $\Delta\chi^2_{\rm min}$ [Eq.~\eqref{eq:chi-squared}], and the logarithmic Bayesian evidence ratio, $\Delta \ln B$ [Eq.~\eqref{eq:bayes-ratio}], are reported at the end of the table.}
\label{tab:model2}

 \end{table}
 
\subsection{Model 1: $ w(a) = w_0\big[1 + \sin(1-a)\big]$}

Table~\ref{tab:model1} and Fig.~\ref{fig:model1} summarize the observational constraints on this dark energy model from the three combined analyses described above. Regardless of the choice of SNe sample, the best-fitting $w(a)$ begins in the phantom regime, crosses the phantom divide at $a \simeq 0.7$, and remains in the quintessence regime thereafter.

The mean equation of state parameter, $w_0$, deviates from $-1$ at more than $5\sigma$ significance for each dataset combination. However, caution is required when interpreting these results. Since $w_0$ determines both the present-day value and the phantom crossing of $w(a)$, our constraints on these features are not independent. Furthermore, $\Lambda$CDM is not nested within our dark energy models, so the deviation of $w_0$ from $-1$ does not represent direct evidence for dynamical dark energy. 
Nevertheless, the preferred present value and phantom crossing are consistent across the dataset combinations and agree well with the results from the CPL parametrization.

Fig.~\ref{fig:model1} presents the marginalized 1D and 2D posteriors for the three derived parameters most relevant to cosmological tensions: $\Omega_m$, $H_0$, and $S_8$. We observe a significant correlation in the $w_0$–$\Omega_m$ and $w_0$–$H_0$ joint posteriors, in addition to the expected correlation between $\Omega_m$ and $H_0$. These correlations lead to small shifts in the mean $\Omega_m$ and $H_0$ from their \textit{Planck} 2018 $\Lambda$CDM values, with $H_0$ increasing when using PantheonPlus and decreasing when using the DESY5 or Union3 samples.

To evaluate the improvement of this model over $\Lambda$CDM and the CPL parametrization, we use the metrics described in the previous section: the change in the minimum chi-squared, $\Delta\chi^2_{\rm min}$, and the logarithmic Bayes ratio, $\Delta \ln B$. The values of these metrics for Model~\hyperref[eq:model1]{1} are consistent with the fact that this model reproduces the $w(a)$ preferred by the CPL parametrization using one fewer parameter: the chi-squared difference from CPL is near zero, while the Bayesian evidence ratios are between $1$–$2$ (Table~\ref{tab:preference-stats}). Hence, the model provides a similar improvement over $\Lambda$CDM as does the CPL parametrization.

\begin{table}
\scalebox{0.75}{ 
\begin{tabular} { l  c c c}
\noalign{\vskip 3pt}\hline\noalign{\vskip 1.5pt}\hline\noalign{\vskip 5pt}
 \multicolumn{1}{c}{\bf Parameter} &  \multicolumn{1}{c}{\bf PantheonPlus} &  \multicolumn{1}{c}{\bf DESY5} &  \multicolumn{1}{c}{\bf Union3}\\\noalign{\vskip 3pt}\hline\noalign{\vskip 1.5pt}\hline\noalign{\vskip 5pt}
{$\Omega_\mathrm{c} h^2$} & $0.11969\pm 0.00072        $ & $0.11937\pm 0.00072        $ & $0.11925\pm 0.00078        $\\

{$\Omega_\mathrm{b} h^2$} & $0.02240\pm 0.00013        $ & $0.02242\pm 0.00013        $ & $0.02243\pm 0.00013        $\\

{$100\theta_\mathrm{MC}$} & $1.04097\pm 0.00028        $ & $1.04099\pm 0.00028        $ & $1.04101\pm 0.00028        $\\

{$\tau_\mathrm{reio}$} & $0.0538\pm 0.0070          $ & $0.0547\pm 0.0069          $ & $0.0553\pm 0.0071          $\\

{$n_\mathrm{s}   $} & $0.9662\pm 0.0034          $ & $0.9669\pm 0.0035          $ & $0.9674\pm 0.0035          $\\

{$\log(10^{10} A_\mathrm{s})$} & $3.042\pm 0.013            $ & $3.044\pm 0.013            $ & $3.045\pm 0.013            $\\

{$w_0            $} & $-0.862\pm 0.023           $ & $-0.839\pm 0.021           $ & $-0.829^{+0.028}_{-0.025}  $\\

{$\Omega_\mathrm{m}$} & $0.3123\pm 0.0052          $ & $0.3169\pm 0.0053          $ & $0.3192\pm 0.0062          $\\

{$\sigma_8       $} & $0.8157\pm 0.0086          $ & $0.8088\pm 0.0082          $ & $0.8059\pm 0.0094          $\\

{$S_8            $} & $0.8321\pm 0.0075          $ & $0.8313\pm 0.0076          $ & $0.8312\pm 0.0077          $\\

{$H_0            $} & $67.62\pm 0.58             $ & $67.05\pm 0.56             $ & $66.78\pm 0.68             $\\

{$r_\mathrm{drag}$} & $147.15\pm 0.20            $ & $147.21\pm 0.20            $ & $147.23\pm 0.21            $\\
\noalign{\vskip 3pt}\hline\noalign{\vskip 3pt}
 $\Delta\chi^2_{\rm min}\:(\Delta\ln B)\:{\Lambda\rm CDM}$ & $-$10.6 (1.2) & $-$19.4 (6.0) & $-$14.0 (3.7) \\ 
 $\Delta\chi^2_{\rm min}\:(\Delta\ln B)\:\mathrm{CPL}$ & $-$1.5 (3.0) & $-$0.5 (2.4) & $-$0.4 (1.6) \\ 
 \noalign{\vskip 3pt}\hline\noalign{\vskip 3pt} 
 \end{tabular}
}
\caption{\textbf{Model 3}. Constraints (68\% CL) on the seven free parameters listed in Table~\ref{table:priors}, alongside relevant derived parameters, using the combination of our baseline dataset with each of the three Type Ia SNe samples (PantheonPlus, DESY5, and Union3) separately. The minimum chi-squared difference, $\Delta\chi^2_{\rm min}$ [Eq.~\eqref{eq:chi-squared}], and the logarithmic Bayesian evidence ratio, $\Delta \ln B$ [Eq.~\eqref{eq:bayes-ratio}], are reported at the end of the table.}
\label{tab:model3}

\end{table}

\subsection{Model 2: $ w(a) = w_0 \left[1 +  \frac{1-a}{a^2 + (1-a)^2}  \right]$}

Table~\ref{tab:model2} and Fig.~\ref{fig:model2} summarize the observational constraints. For all three SNe catalogs, the best-fitting $w(a)$ has a phantom crossing at $a \simeq 0.7$, consistent with the previous model. 
Again, the posterior distribution of the equation of state parameter shows a preference for $w_0 > -1$ at greater than $5\sigma$ significance.

The posterior distributions of $\Omega_m$, $H_0$, and $S_8$ are consistent with the \textit{Planck} 2018 $\Lambda$CDM results~\cite{Planck:2018vyg} within 68\% CL only when PantheonPlus or DESY5 are used. The mean values obtained for $H_0$ and $\Omega_m$ with the Union3 dataset deviate from the mean \textit{Planck} 2018 $\Lambda$CDM parameters at approximately $2\sigma$ significance. This is likely related to the geometrical degeneracy in the $w_0$–$\Omega_m$ and $w_0$–$H_0$ planes and the differing $\Omega_m$ preferences of the SNe samples. 
The highest mean value of the Hubble constant, $H_0 = 67.07 \pm 0.59$ km\,s$^{-1}$\,Mpc$^{-1}$ (68\% CL), is obtained using the PantheonPlus dataset and is slightly below the canonical \textit{Planck} 2018 $\Lambda$CDM value.

The model preference statistics at the end of Table~\ref{tab:model2} strongly depend on the choice of SNe sample. While both the chi-squared and Bayesian evidence favor this model over $\Lambda$CDM when DESY5 or Union3 data are used, PantheonPlus yields $\Delta\chi^2_{\rm min} \simeq 5$ and $\Delta\ln B \simeq -6.5$, indicating strong support for $\Lambda$CDM. This latter dataset also strongly disfavors Model~\hyperref[eq:model2]{2} when compared to the CPL parametrization—the only case where this occurs (Table~\ref{tab:preference-stats} and Fig.~\ref{fig:chi2-lnB})—while the other two dataset combinations remain mostly agnostic. These weakened preferences are perhaps due to this equation of state’s unusual shape. To match the phantom crossing preferred by the CPL parametrization, it must deviate from CPL near $a = 1$ at approximately $2\sigma$ significance (see the lower panel of Fig.~\ref{fig:model2}).

\begin{table}
\scalebox{0.75}{ 
\begin{tabular} { l  c c c}
\noalign{\vskip 3pt}\hline\noalign{\vskip 1.5pt}\hline\noalign{\vskip 5pt}
 \multicolumn{1}{c}{\bf Parameter} &  \multicolumn{1}{c}{\bf PantheonPlus} &  \multicolumn{1}{c}{\bf DESY5} &  \multicolumn{1}{c}{\bf Union3}\\\noalign{\vskip 3pt}\hline\noalign{\vskip 1.5pt}\hline\noalign{\vskip 5pt}
{$\Omega_\mathrm{c} h^2$} & $0.12008\pm 0.00072        $ & $0.11984\pm 0.00074        $ & $0.11973\pm 0.00075        $\\

{$\Omega_\mathrm{b} h^2$} & $0.02237\pm 0.00013        $ & $0.02239\pm 0.00013        $ & $0.02240\pm 0.00013        $\\

{$100\theta_\mathrm{MC}$} & $1.04091\pm 0.00028        $ & $1.04094\pm 0.00029        $ & $1.04095\pm 0.00028        $\\

{$\tau_\mathrm{reio}$} & $0.0524\pm 0.0068          $ & $0.0532\pm 0.0069          $ & $0.0539\pm 0.0070          $\\

{$n_\mathrm{s}   $} & $0.9652\pm 0.0035          $ & $0.9657\pm 0.0034          $ & $0.9662\pm 0.0034          $\\

{$\log(10^{10} A_\mathrm{s})$} & $3.039\pm 0.012            $ & $3.041\pm 0.013            $ & $3.042\pm 0.013            $\\

{$w_0            $} & $-0.940\pm 0.026           $ & $-0.918\pm 0.024           $ & $-0.906\pm 0.031           $\\

{$\Omega_\mathrm{m}$} & $0.3128\pm 0.0055          $ & $0.3169\pm 0.0051          $ & $0.3194\pm 0.0064          $\\

{$\sigma_8       $} & $0.8179\pm 0.0084          $ & $0.8124\pm 0.0083          $ & $0.8093\pm 0.0097          $\\

{$S_8            $} & $0.8350\pm 0.0074          $ & $0.8349\pm 0.0077          $ & $0.8350\pm 0.0075          $\\

{$H_0            $} & $67.65\pm 0.60             $ & $67.15\pm 0.54             $ & $66.87\pm 0.71             $\\

{$r_\mathrm{drag}$} & $147.08\pm 0.20            $ & $147.13\pm 0.20            $ & $147.14\pm 0.20            $\\
\noalign{\vskip 3pt}\hline\noalign{\vskip 3pt}
 $\Delta\chi^2_{\rm min}\:(\Delta\ln B)\:{\Lambda\rm CDM}$ & $-$12.6 (2.7) & $-$15.8 (3.9) & $-$12.7 (2.7) \\ 
 $\Delta\chi^2_{\rm min}\:(\Delta\ln B)\:\mathrm{CPL}$ & $-$3.5 (4.6) & 3.1 (0.3) & 0.9 (0.6) \\ 
 \noalign{\vskip 3pt}\hline\noalign{\vskip 3pt} 
 \end{tabular}
}

\caption{\textbf{Model 4}. Constraints (68\% CL) on the seven free parameters listed in Table~\ref{table:priors}, alongside relevant derived parameters, obtained using the combination of our baseline dataset with each of the three Type Ia SNe samples (PantheonPlus, DESY5, and Union3) separately. The minimum chi-squared difference, $\Delta\chi^2_{\rm min}$ [Eq.~\eqref{eq:chi-squared}], and the logarithmic Bayesian evidence ratio, $\Delta \ln B$ [Eq.~\eqref{eq:bayes-ratio}], are reported at the end of the table.}
\label{tab:model4}

\end{table}

\begin{figure}
    \centering
    \includegraphics[width=\linewidth]{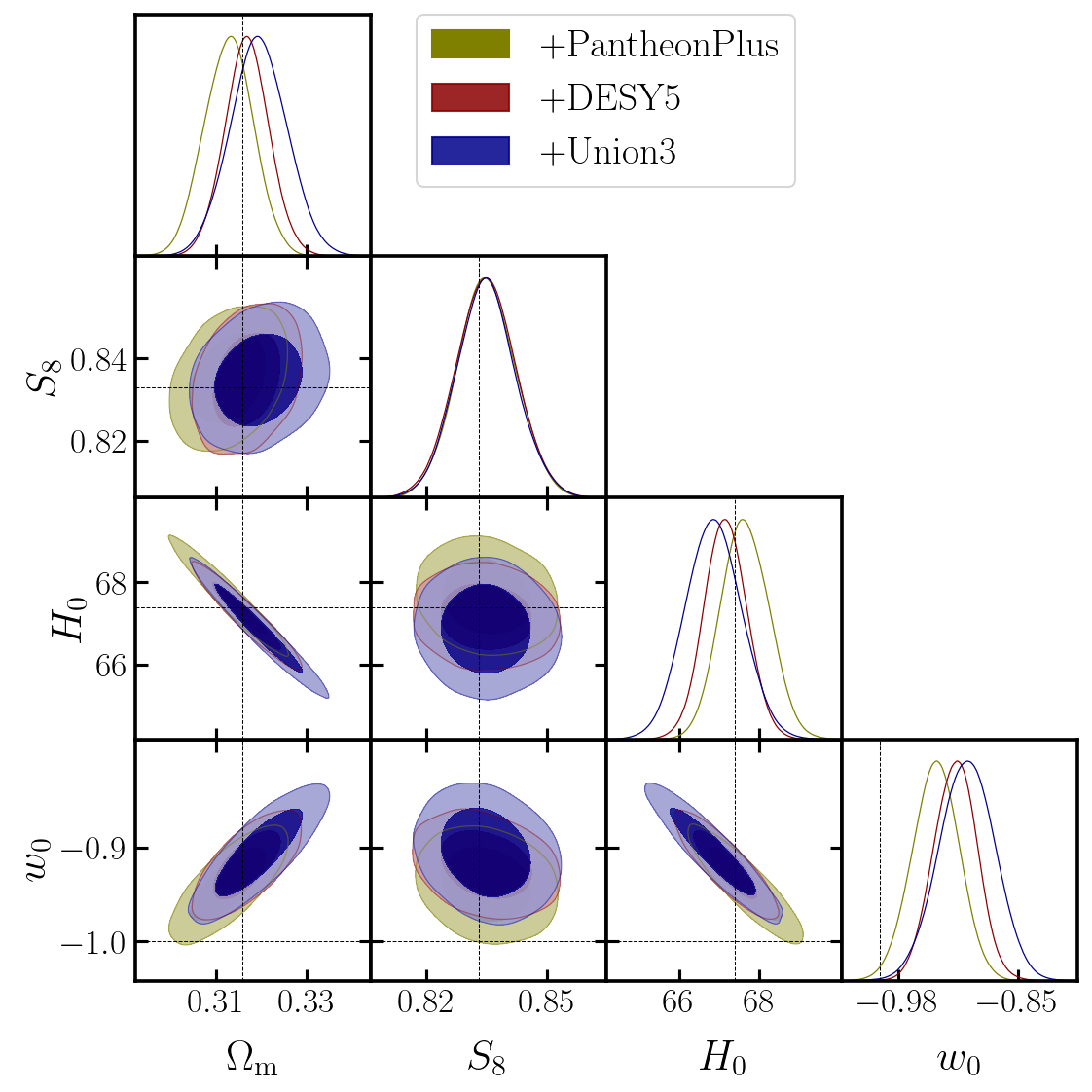}
    \includegraphics[width=\linewidth]{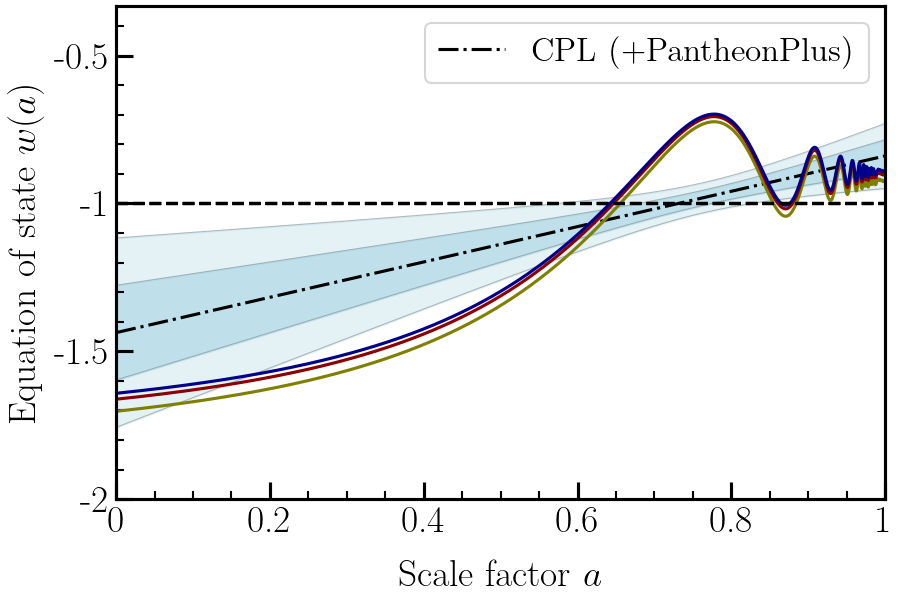}
    \caption{\textbf{Model 4}. \textit{Above:} Posterior distributions and $1\sigma$, $2\sigma$ contours for three key cosmological parameters—the dark energy equation of state parameter $w_0$, the Hubble constant $H_0$, and the matter fluctuation parameter $S_8$—obtained from the combination of our baseline dataset with each of the three Type Ia SNe samples, separately, as indicated in the legend. The dashed lines represent the mean $\Lambda$CDM parameter values from \textit{Planck} 2018~\cite{Planck:2018vyg}.  
    \textit{Below:} The best-fitting $w(a)$ for the three dataset combinations shown in the legend above. The dash-dot line shows the mean $w(a)$ obtained using the CPL parametrization with the PantheonPlus combination, while the light blue bands indicate the corresponding $1\sigma$ and $2\sigma$ intervals.}
    \label{fig:model4}
\end{figure}

\begin{figure}
    \centering
    \includegraphics[width=\linewidth]{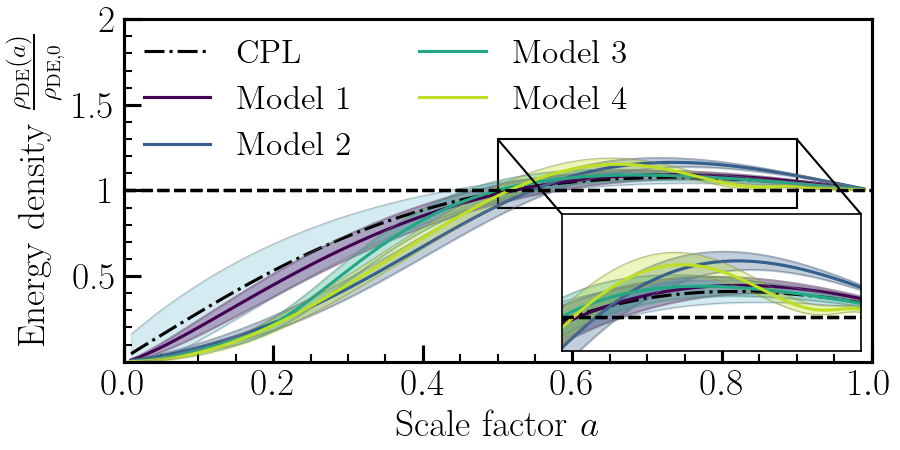}
    \caption{The mean energy density, $\rho_{\rm DE}(a)$, and $1\sigma$ bands for the CPL parametrization (light blue) and Models~\hyperref[eq:model1]{1}--\hyperref[eq:model4]{4} (colors indicated in the legend), obtained using the dataset combination involving PantheonPlus SNe. Model~\hyperref[eq:model4]{4} is notable for its well-defined peak near $a \simeq 2/3$.}
    \label{fig:rho-de}
\end{figure}

\subsection{Model 3: $w(a) = w_0 \left[1 - a \sin \left(\frac{1}{a} \right)+ \sin1 \right]$}

Table~\ref{tab:model3} and Fig.~\ref{fig:model3} summarize the observational constraints. 
This $w(a)$ oscillates with decreasing frequency and increasing amplitude over cosmic time. Most of its oscillations occur within $a \lesssim 0.2$, after which the equation of state evolves monotonically. 
The best-fitting $w(a)$ closely matches the CPL parametrization for $a \gtrsim 0.5$, suggesting that the early-time oscillations may not significantly affect the preferred value of $w_0$.
Similar to the previous models, there is a phantom crossing at $a \simeq 0.7$, and the present value of $w(a)$ deviates from $-1$ at high significance for each dataset combination. 

The posterior distributions of the three derived parameters in Fig.~\ref{fig:model3} ($\Omega_m$, $H_0$, and $S_8$) show qualitative similarities to those of Model~\hyperref[eq:model1]{1}. The $w_0$ parameter remains strongly correlated with $\Omega_m$ and $H_0$, and the mean values of all three parameters are slightly shifted but remain well within $1\sigma$ of the \textit{Planck} 2018 $\Lambda$CDM results~\cite{Planck:2018vyg}.

The model comparison metrics for this model are also qualitatively similar to those of Model~\hyperref[eq:model1]{1} (Table~\ref{tab:preference-stats} and Fig.~\ref{fig:chi2-lnB}), providing a similar improvement over $\Lambda$CDM to the CPL parametrization. However, this model improves over Model~\hyperref[eq:model1]{1} in fitting the dataset involving PantheonPlus. Using this dataset, the Bayesian evidence for Model~\hyperref[eq:model3]{3} compared to CPL becomes moderate ($\Delta\ln B \simeq 3$). Although this improvement could arise from the differences between Models~\hyperref[eq:model1]{1} and~\hyperref[eq:model3]{3} during late times (for example, the slightly earlier phantom crossing of Model~\hyperref[eq:model3]{3}), an intriguing possibility is that the markedly different early-time behavior of Model~\hyperref[eq:model3]{3} has beneficial downstream effects for fitting the SNe distance measurements.

\subsection{Model 4: $ w(a) = w_0 \left[1 + (1-a)\sin\left(\frac{1}{1-a}\right) \right]$}

Table~\ref{tab:model4} and Fig.~\ref{fig:model4} summarize the observational constraints. The $w(a)$ of this model oscillates with \textit{increasing} frequency and an amplitude that \textit{decreases} over cosmic time, reaching zero at the present day.\footnote{After the present epoch ($a = 1$), the equation of state follows a time-reversed evolution, with oscillations decreasing in frequency and increasing in amplitude. Thus, while $w(a)$ can be extended indefinitely into the future without diverging, the special role assigned to the present day is unjustified. This model should therefore be viewed as a phenomenological framework that allows for present-day oscillations within a generally increasing trend.}  
The late-time behavior of the best-fitting $w(a)$ is remarkably consistent across the dataset combinations, featuring a phantom crossing at $a \simeq 2/3$ and a present value of $w(a)$ that differs from $-1$ at approximately $3\sigma$ significance. This represents the smallest deviation from $-1$ among the four models considered. Although one could interpret this as suggestive evidence for present-day oscillations near the cosmological constant, such oscillations are not the only distinctive feature of this model. For instance, the equation of state also has a well-defined peak at $a \simeq 0.78$, regardless of the value of $w_0$ (Fig.~\ref{fig:gradient}).  
For the best-fitting values of $w_0$, this peak rises far above the $2\sigma$ contour of the CPL parametrization (Fig.~\ref{fig:model4}) and leads to a similar deviation in energy density, $\rho_{\rm DE}(a)$ (Fig.~\ref{fig:rho-de}).

Even though this model has a mean $w_0$ much closer to $-1$ than the previous three models, the posterior distributions of $\Omega_m$, $H_0$, and $S_8$ are similar. The Hubble constant is again slightly higher when PantheonPlus is used, but for every dataset combination considered, the mean values of these derived parameters are consistent with \textit{Planck} 2018 $\Lambda$CDM~\cite{Planck:2018vyg} to within 68\% CL.  

The model comparison metrics in Table~\ref{tab:model4} show that Model~\hyperref[eq:model4]{4} is favored over $\Lambda$CDM by both of our statistical metrics for every dataset combination considered. In particular, this model provides a significantly better fit to the dataset involving PantheonPlus than Models~\hyperref[eq:model1]{1}--\hyperref[eq:model3]{3} and the CPL parametrization (Table~\ref{tab:preference-stats} and Fig.~\ref{fig:chi2-lnB}), achieving $\Delta\chi^2_{\rm min} \simeq -3.5$ and $\Delta \ln B \simeq 4.6$ compared to CPL. However, the model also provides a significantly worse fit to the dataset involving DESY5 than Models~\hyperref[eq:model1]{1} and~\hyperref[eq:model3]{3}, although its performance is comparable to CPL in terms of Bayesian evidence. Every model considered, including Model~\hyperref[eq:model2]{2}, provides a comparable fit to the Union3 dataset.

Regarding the preference of the PantheonPlus dataset for Model~\hyperref[eq:model4]{4}, and the preference of the DESY5 dataset for models with more linear late-time evolution, one explanation lies in the behavior of $w(a)$ near $a \simeq 2/3$ (redshift $z \simeq 0.5$). Recent non-parametric reconstructions~\cite{Bansal:2025ipo,Ormondroyd:2025exu,Ormondroyd:2025iaf,Ormondroyd:2025iaf,Mukherjee:2025fkf,Berti:2025phi,Gu:2025xie} have found deviations from $\Lambda$CDM in various cosmological functions, including $w(a)$, near this characteristic time. In particular, Refs.~\cite{Ormondroyd:2025exu,Ormondroyd:2025iaf,Ormondroyd:2025iaf} demonstrated that the deviation in $w(a)$ near $a \simeq 2/3$ is more pronounced when the DESI BAO data are combined with PantheonPlus than with the DESY5 sample. Our results support these findings within the parametric approach. From a more fundamental perspective, this qualitative behavior could result from, e.g., a coupling between dark energy and dark matter~\cite{Smith:2024ibv,Khoury:2025txd}.

\section{\label{sec:summary}Summary and Conclusions}

\begin{figure*}
    \centering
    \includegraphics[width=\linewidth]{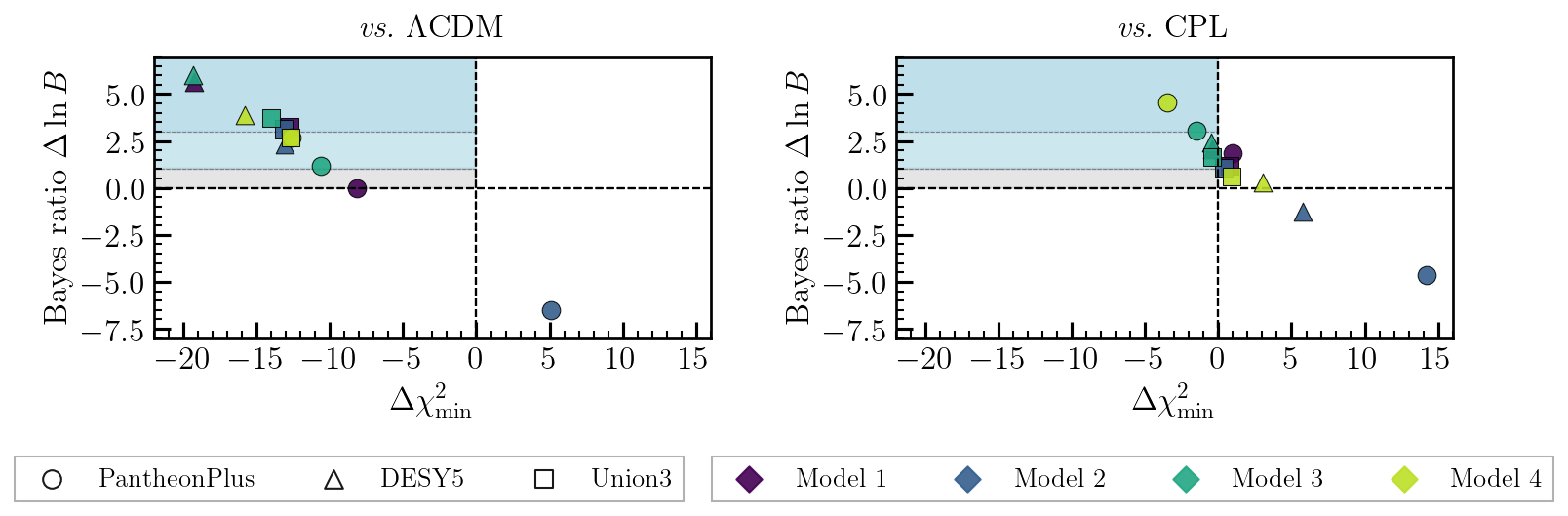}
    \caption{A summary of the datasets’ model preferences, as quantified by the change in the minimum chi-squared, $\Delta\chi^2_{\rm min}$ [Eq.~\eqref{eq:chi-squared}], and the logarithmic Bayes ratio, $\Delta \ln B$ [Eq.~\eqref{eq:bayes-ratio}], for each one-parameter model in Eqs.~\eqref{eq:model1}--\eqref{eq:model4}. The dataset combinations are defined by the Type Ia SNe sample (PantheonPlus, DESY5, or Union3) used alongside our baseline (Section~\ref{sec:data}).  
    The top-left quadrants indicate preference for our models, and the shaded bands demarcate the revised Jeffreys’ scale (Section~\ref{sec:preference-stats}). Compared to $\Lambda$CDM (\textit{Left}), all models except Model~\hyperref[eq:model2]{2} with PantheonPlus SNe are significantly favored by one or both statistics. Versus the CPL parametrization (\textit{Right}), Model~\hyperref[eq:model4]{4} combined with PantheonPlus SNe provides the most significant improvement.}
    \label{fig:chi2-lnB}
\end{figure*}

\begin{table}[]
    \centering
    \scalebox{0.85}{ 
    \begin{tabular}{ l c c c}
        \noalign{\vskip 3pt}\hline\noalign{\vskip 1.5pt}\hline\noalign{\vskip 5pt}
         \multicolumn{1}{c}{\bf Parameter} &  \multicolumn{1}{c}{\bf PantheonPlus} &  \multicolumn{1}{c}{\bf DESY5} &  \multicolumn{1}{c}{\bf Union3}\\\noalign{\vskip 3pt}\hline\noalign{\vskip 1.5pt}\hline
        \noalign{\vskip 2pt}
        \multicolumn{4}{c}{CPL} \\
        \noalign{\vskip 2pt}
        \hline
        \noalign{\vskip 2pt}
        $\Delta\chi^2_{\rm min}\:(\Delta\ln B)\:{\Lambda\rm CDM}$ & $-$9.2 ($-$1.9) & $-$18.8 (3.6) & $-$13.6 (2.1)
        \\
        \noalign{\vskip 2pt}
        \hline
        \noalign{\vskip 2pt}
        \multicolumn{4}{c}{Model 1} \\
        \noalign{\vskip 2pt}
        \hline
        \noalign{\vskip 2pt}
        $\Delta\chi^2_{\rm min}\:(\Delta\ln B)\:{\Lambda\rm CDM}$ & $-$8.2 ($-$0.0) & $-$19.3 (5.6) & $-$12.8 (3.3) \\ 
        $\Delta\chi^2_{\rm min}\:(\Delta\ln B)\:\mathrm{CPL}$ & 1.0 (1.8) & $-$0.4 (2.1) & 0.8 (1.2) \\ 
        \noalign{\vskip 2pt}
        \hline
        \noalign{\vskip 2pt}
        \multicolumn{4}{c}{Model 2} \\
        \noalign{\vskip 2pt}
        \hline
        \noalign{\vskip 2pt}
        $\Delta\chi^2_{\rm min}\:(\Delta\ln B)\:{\Lambda\rm CDM}$ & 5.1 ($-$6.5) & $-$13.1 (2.3) & $-$13.2 (3.2) \\ 
        $\Delta\chi^2_{\rm min}\:(\Delta\ln B)\:\mathrm{CPL}$ & 14.2 ($-$4.6) & 5.7 ($-$1.3) & 0.4 (1.1) \\ 
        \noalign{\vskip 2pt}
        \hline
        \noalign{\vskip 2pt} 
        \multicolumn{4}{c}{Model 3} \\
        \noalign{\vskip 2pt}
        \hline
        \noalign{\vskip 2pt}
        $\Delta\chi^2_{\rm min}\:(\Delta\ln B)\:{\Lambda\rm CDM}$ & $-$10.6 (1.2) & $-$19.4 (6.0) & $-$14.0 (3.7) \\ 
        $\Delta\chi^2_{\rm min}\:(\Delta\ln B)\:\mathrm{CPL}$ & $-$1.5 (3.0) & $-$0.5 (2.4) & $-$0.4 (1.6) \\ 
        \noalign{\vskip 2pt}
        \hline
        \noalign{\vskip 2pt} 
        \multicolumn{4}{c}{Model 4} \\
        \noalign{\vskip 2pt}
        \hline
        \noalign{\vskip 2pt}
        $\Delta\chi^2_{\rm min}\:(\Delta\ln B)\:{\Lambda\rm CDM}$ & $-$12.6 (2.7) & $-$15.8 (3.9) & $-$12.7 (2.7) \\ 
        $\Delta\chi^2_{\rm min}\:(\Delta\ln B)\:\mathrm{CPL}$ & $-$3.5 (4.6) & 3.1 (0.3) & 0.9 (0.6) \\ 
        \noalign{\vskip 2pt}
        \hline
    \end{tabular}
    }
    \caption{A summary of model preference statistics—the minimum chi-squared difference, $\Delta\chi^2_{\rm min}$ [Eq.~\eqref{eq:chi-squared}], and the logarithmic Bayesian evidence ratio, $\Delta \ln B$ [Eq.~\eqref{eq:bayes-ratio}]—for the CPL parametrization and the one-parameter models considered in this work. A preference for Models~\hyperref[eq:model1]{1}--\hyperref[eq:model4]{4} is theoretically indicated by $\Delta\chi^2_{\rm min} > 0$ and $\Delta \ln B > 0$, while $\Delta\chi^2_{\rm min} < 0$ and $\Delta \ln B < 0$ indicate a preference for the reference models. More precise benchmarks for $\Delta \ln B$ are provided by the revised Jeffreys’ scale (Section~\ref{sec:preference-stats}).}
    \label{tab:preference-stats}
\end{table}

In this article, we have examined four dynamical dark energy models [Eqs.~\eqref{eq:model1}--\eqref{eq:model4}], each with a single free parameter, $w_0$, that controls both the present-day value and the shape of the dark energy equation of state, $w(a)$. All of these equations of state remain well-defined and contain some oscillations over the full history of the universe, in contrast to the commonly assumed CPL parametrization, which diverges in the infinite future ($a \rightarrow \infty$) and cannot capture deviations from linearity. Unlike the CPL parametrization, which arises from the series expansion of a general $w(a)$ to linear order, our models are purely phenomenological, representing minimal extensions of the $\Lambda$CDM model that allow for a dark energy equation of state that broadly increases from the beginning of the universe until the present day. Our main objective was to assess whether any of these models can provide a better fit to modern cosmological data, as measured by the chi-squared statistic and Bayesian evidence, than the CPL parametrization.

Among the four models considered, only one evolves monotonically over the time domain $a \in [0,1]$ that is relevant to our analysis. One other model has a single “oscillation,” decreasing until $a \simeq 0.3$ and increasing thereafter. The remaining two models exhibit rapid oscillations either toward the beginning of the universe ($a = 0$) or the present day ($a = 1$). The latter Model~\hyperref[eq:model4]{4} is particularly noteworthy, as it simultaneously captures the increasing trend suggested by previous studies (e.g., those by DESI~\cite{DESI:2024mwx,DESI:2024aqx,DESI:2025zgx,Lodha:2025qbg}) and the late-time oscillations identified in non-parametric reconstructions~\cite{Zhao:2017cud,Wang:2018fng,Zhang:2019jsu,Escamilla:2021uoj,Escamilla:2024fzq,Jiang:2024xnu,Ye:2024ywg,Ormondroyd:2025exu,Ormondroyd:2025iaf,Berti:2025phi,Gu:2025xie}. To our knowledge, this model has not been previously considered in the literature.

These models were analyzed using a combination of CMB data from \textit{Planck}, lensing reconstruction from \textit{Planck} and ACT-DR6, BAO measurements from DESI-DR2, and three separate Type Ia supernovae samples: PantheonPlus, DESY5, and Union3. The resulting parameter constraints are presented in Tables~\ref{tab:model1}--\ref{tab:model4} and Figs.~\ref{fig:model1}--\ref{fig:model4}. Each best-fitting $w(a)$ has a present-day value greater than $-1$ at several standard deviations and a phantom crossing near $a \simeq 0.6$--$0.7$.  
This behavior is qualitatively consistent with both the CPL parametrization and non-parametric reconstruction results from DESI~\cite{DESI:2024mwx,DESI:2024aqx,DESI:2025zgx,Lodha:2025qbg}.

Our main results regarding model preferences---measured by the minimum chi-squared difference, $\Delta\chi^2_{\rm min}$ [Eq.~\eqref{eq:chi-squared}], and logarithmic Bayes ratio, $\Delta\ln B$ [Eq.~\eqref{eq:bayes-ratio}]---for our one-parameter models compared to $\Lambda$CDM and the CPL parametrization are shown in Fig.~\ref{fig:chi2-lnB} and Table~\ref{tab:preference-stats}. These results can be summarized as follows:
\begin{itemize}
    \item All datasets significantly prefer the one-parameter models over $\Lambda$CDM according to the chi-squared statistic, with the exception of Model~\hyperref[eq:model2]{2} and the dataset involving PantheonPlus, which is significantly disfavored. The DESY5 dataset provides the highest Bayesian evidence ($\Delta\ln B \simeq 5$) for Models~\hyperref[eq:model1]{1} and~\hyperref[eq:model3]{3}, whereas the Union3 dataset provides similar levels of support for each model.
    
    \item Model~\hyperref[eq:model4]{4} with PantheonPlus is significantly favored over the CPL parametrization by the Bayesian evidence, with $\Delta\ln B \simeq 4.5$. 
    Using the same dataset, Model~\hyperref[eq:model3]{3} is moderately favored ($\Delta\ln B \simeq 3$). All other datasets are either indecisive or favor the CPL parametrization over the one-parameter models.
    
    \item The strong preference for Model~\hyperref[eq:model4]{4} from PantheonPlus is possibly explained by this model’s $w(a)$ having a well-defined peak near $a \simeq 0.78$, leading to a similar peak in $\rho_{\rm DE}(a)$ near $a \simeq 2/3$ (Fig.~\ref{fig:rho-de}). 
    Such features were also found near this scale factor in non-parametric reconstructions of $w(a)$ using PantheonPlus SNe (less so with DESY5)~\cite{Ormondroyd:2025exu,Ormondroyd:2025iaf}. This qualitative behavior could result from, e.g., a coupling between dark energy and dark matter~\cite{Smith:2024ibv,Khoury:2025txd}.
\end{itemize}

In conclusion, we have identified several one-parameter models of dark energy that are preferred over $\Lambda$CDM and the CPL parametrization by standard statistical metrics, depending on the SNe catalog used in the analysis. Our results add to the growing collection of hints for deviations from $\Lambda$CDM near $a \simeq 2/3$ ($z \simeq 0.5$)~\cite{Bansal:2025ipo,Ormondroyd:2025exu,Ormondroyd:2025iaf,Ormondroyd:2025iaf,Mukherjee:2025fkf,Berti:2025phi,Gu:2025xie}, motivating further studies on dark energy models with late-time oscillations.

\textit{Added note:} Recent investigations into the photometric calibration of Type Ia supernova samples, most notably the findings of the ``Dovekie'' analysis \cite{Popovic:2025glk, DES:2025sig}, have identified systematic biases affecting both the DESY5 and Pantheon+ catalogs. While a full re-analysis using a recalibrated Pantheon+ likelihood is currently restricted by data availability, we can estimate the potential impact on our results by extrapolating from recent studies. Estimates based on \cite{Popovic:2025glk, DES:2025sig} suggest that recalibration typically shifts the equation of state parameters $w_0$ and $w_a$ toward their $\Lambda$CDM values (e.g., $\Delta w_0 \sim 0.05$ and $\Delta \Omega_m \sim -0.02$). 
However, in addition to supernovae photometric calibration, the impact of host-galaxy corrections has received renewed attention. The recent Union3.1 compilation provides a consistent re-evaluation of host galaxy masses across $\sim 2000$ Type Ia supernovae, successfully resolving historical low-redshift discrepancies~\cite{Hoyt:2026fve}. While these improvements systematically shift the baseline matter density by $\Delta\Omega_m \sim 0.01$, Ref.~\cite{Hoyt:2026fve} does not incorporate the Dovekie photometric offsets derived by Ref.~\cite{Popovic:2021yuo}, which can introduce separate systematic variations of $\Delta\Omega_m\sim 0.05$. Until the interplay between these independent corrections is fully understood, studies incorporating supernova data to evaluate the significance of dynamical dark energy should be treated with appropriate caution.

\section{Acknowledgments}
We sincerely thank the referee for their valuable comments which improved the manuscript.
We thank William Giar\`e for the interesting and useful discussions and for his contribution in developing the Cobaya wrapper for the MCEvidence.
SP acknowledges the financial support from the Department of Science and Technology (DST), Govt. of India under the Scheme  ``Fund for Improvement of S\&T Infrastructure (FIST)'' [File No. SR/FST/MS-I/2019/41]. EDV is supported by a Royal Society Dorothy Hodgkin Research Fellowship.
This article is based upon work from the COST Action CA21136 ``Addressing observational tensions in cosmology with systematics and fundamental physics'' (CosmoVerse), supported by COST (European Cooperation in Science and Technology). We acknowledge IT Services at The University of Sheffield for the provision of services for High
Performance Computing.

\appendix
\section{Dataset Consistency}
\appendix

\begin{figure}
    \centering
    \includegraphics[width=\linewidth]{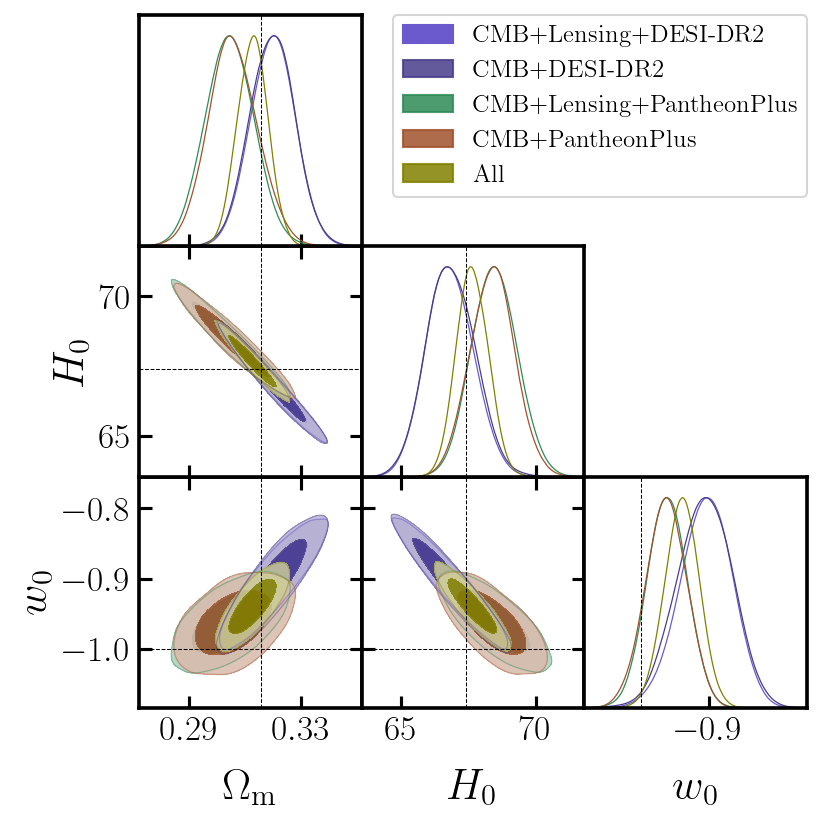}
    \caption{\textbf{Model 4}. {Posterior distributions obtained from subsets of the full dataset combination (CMB data from \textit{Planck}, lensing reconstruction from \textit{Planck} and
    ACT-DR6, BAO measurements from DESI-DR2, and
    Type Ia supernovae from PantheonPlus) that provided our most statistically significant results. When analyzed within Model 4, all datasets are consistent with one another to within 1$\sigma$.}}
    \label{fig:dataset-consistency}
\end{figure}
\begin{table}[]
    \centering
    \scalebox{0.79}{ 
    \begin{tabular}{ l c c}
        \noalign{\vskip 3pt}\hline\noalign{\vskip 1.5pt}\hline\noalign{\vskip 5pt}
         \multicolumn{1}{c}{\bf Dataset} &  \multicolumn{1}{c}{$\Delta\chi^2_{\rm min}\:(\Delta\ln B)\:\Lambda\mathrm{CDM}$} &  \multicolumn{1}{c}{$\Delta\chi^2_{\rm min}\:(\Delta\ln B)\:\mathrm{CPL}$}\\\noalign{\vskip 3pt}\hline\noalign{\vskip 1.5pt}\hline
        \noalign{\vskip 2pt}
        CMB+Lensing+DESI-DR2 & $-$8.4 (1.2) & 0.2 (1.2) \\
        CMB+DESI-DR2 & $-$9.3 (1.3) & 1.0 (1.3) \\
        CMB+Lensing+PantheonPlus & $-$4.5 ($-$1.2) & $-$2.4 (3.2) \\
        CMB+PantheonPlus & $-$3.9 ($-$1.3) & $-$3.6 (3.0) \\
        All & $-$12.6 (2.7) & $-$3.5 (4.6)  \\
        \noalign{\vskip 2pt}
        \hline
    \end{tabular}
    }
    \caption{\textbf{Model 4.} {Model preference statistics from subsets of the full dataset combination (CMB data from Planck,
    lensing reconstruction from Planck and ACT-DR6, BAO mea-
    surements from DESI-DR2, and Type Ia supernovae from
    PantheonPlus) that provided our most statistically significant
    results. The Bayesian preference for Model 4 over CPL arises primarily from PantheonPlus.}}
    \label{tab:subsets}
\end{table}
In this appendix, we demonstrate the consistency of the datasets combined to obtain our results for Model 4 and its most statistically significant dataset combination: CMB data from \textit{Planck}, lensing reconstruction from \textit{Planck} and ACT-DR6, BAO measurements from DESI-DR2, and
Type Ia supernovae from PantheonPlus. In Figure~\ref{fig:dataset-consistency}, we show select parameter constraints obtained from individual MCMC runs on subsets of the full dataset combination. One can see that, when analyzed within this model, all individual datasets are consistent with one another to within 1$\sigma$. The model comparison statistics for these subsets (Table~\ref{tab:subsets}) confirm that the PantheonPlus catalog is primarily responsible for the Bayesian preference for Model 4 over CPL.
We emphasize that, due to computational cost, this detailed breakdown was performed only for Model~4, which yielded the most statistically significant improvement over CPL and therefore provides a stringent consistency test.\footnote{For completeness, we note that we found similar consistency for this model and subsets of the DESY5 and Union3 dataset combinations. We expect that Models 1--3 would not display less consistency, as these models' phenomenology and model comparison statistics are similar to those of Model 4.}

\bibliography{biblio}

%-----------------------------------------------------------
\end{document}